\documentclass[prd,eqsecnum,tightenlines,11pt]{revtex4}

\usepackage{hyperref}
\usepackage{graphicx}
\usepackage{amsmath,amssymb,amsfonts,amsthm,latexsym,stmaryrd}
\usepackage{marginnote}
\usepackage{color}
\usepackage{soul}

\begin{document}

\title{Axial and polar stability of neutron stars in scalar-tensor theories with disformal coupling}
\author{Hamza Boumaza}
\affiliation{\small Laboratoire de Physique des Particules et Physique Statistique (LPPPS),\\
 Ecole Normale Supérieure-Kouba, B.P. 92, Vieux Kouba, 16050 Algiers, Algeria}

\begin{abstract}
{\hskip 2em}In the present work, we study the radial and non-radial perturbative stability of neutron stars in which the matter is disformally coupled to the metric. First, we derive the gravitational and the fluid equations of the neutron star in a static and spherically symmetric background. Then, we calculate the second-order expansion of the action that describes the dynamics of both axial and polar modes. From the resulting expressions, we derive the conditions to avoid  gradient and ghost instabilities at the center of the star and at spatial infinity. In addition, a numerical analysis is performed to investigate the stability of a particular model with a constant disformal function denoted as $\Lambda$ in the whole space time. We found that the chosen model is stable against the gradient instability in a small range of the constant $\Lambda$.
 \end{abstract}

\maketitle

\section{Introduction:}

 Although the complexity of observing black holes or neutron stars with high precision means that general relativity (GR) has not yet been fully tested in the strong gravity regime, where we can expect to extensions of GR and new physical phenomena, it is so far the most successful  gravitational theory in describing stellar objects and gravitational waves (GWs).  The theory has also successfully explained the existence of compact objects such as neutron stars (NSs) and black holes (BHs), which are the main sources of the recently detected gravitational waves by the LIGO/Virgo Collaboration  \cite{abbott2016gw150914,Abbott_2017,armengol2017neutron,LIGOScientific:2017vwq,abbott2018prospects,abbott2018gw170817,abbott2020gw190814}. These recent measurements have opened a new window for studying the physics of dynamical space-time  in  the strong gravity regime through the study of gravitational wave signals. More precise measurements will also allow us to understand the structure of compact objects, particularly the nature of the matter in the core of neutron stars.  

 Another interesting feature which can be probed by the analysis of GW signals is the presence of deviation from GR in the strong gravity region. In many modified theories of gravity in particular scalar-tensor theories of gravity \cite{fujii2003scalar}, this deviation is a consequence of the presence of new degrees of freedom for example a scalar field. The most general scalar-tensor theory of gravity with second order derivatives that provides  second order equations of motion is Horndeski theory \cite{Horndeski:1974wa}. This theory is generalized to beyond Horndeski \cite{Gleyzes:2014qga} and then to Degenerate-Higher-Order-Scalar-Tensor (DHOST) theories \cite{Langlois:2015cwa}, which contain higher derivatives but without suffering from Ostrogradski's instability \cite{ostrogradsky1850memoires}. Moreover, Horndeski gravity and DHOST theories can be mapped into each other through a conformal-disformal transformations \cite{BenAchour:2016cay}. In these theories, the scalar field plays an important role, which can lead to the late acceleration of the universe \cite{Boumaza:2020klg,Langlois:2018dxi,Langlois:2017dyl} or to deviation from GR modifying black holes \cite{BenAchour:2020wiw,Minamitsuji:2019shy,Motohashi:2019sen} and neutron stars \cite{Boumaza:2021fns,Ogawa:2019gjc,Boumaza:2022abj,Babichev:2016jom,Boumaza:2021lnp,Cisterna:2015yla,Cisterna:2016vdx}. In this paper, we will limit our study to neutron stars.

 An interesting subfamily of scalar tensor theories has been studied intensively in the literature, in which the metric is  coupled to the matter via conformal transformation \cite{Ramazanoglu:2016kul,Yazadjiev:2016pcb,Harada:1998ge},  to describe the neutron star profile. Due to tachyonic instability, it was found that neutron stars, with negative scalar field mass, are spontaneously scalarized for a small range of the parameter in the model \cite{Damour:1996ke,Freire:2012mg}. The phenomenological implications of spontaneous scalarization have been explored for  static neutron stars in many situations, for example: slowly rotating NSs \cite{Sotani:2012eb,Pani:2014jra} where in these papers it is reported that the scalar field can modify the relation between the mass and the moment of inertia. In addition, the amplitude of gravitational waves, sourced by the collapse of  a neutron star  into a black hole,  can be modified by the scalar field \cite{Novak:1997hw}. Other phenomena can be produced in this kind of theory, (see Ref.\cite{Harada:1996wt}).
 
Spontaneous scalarization can also occur when matter is coupled to the metric via disformal transformations \cite{Minamitsuji:2016hkk}, which are the most general transformations of the metric that preserve the causality principle \cite{Bekenstein:1992pj}. The disformal transformation, studied in Ref. \cite{Minamitsuji:2016hkk}, is a generalization of conformal transformations that preserves the mathematical structure of Horndeski theory \cite{Bettoni:2013diz}. Moreover, the disformal coupling of matter has gained considerable interest in recent years, where it has been studied in dark sectors \cite{Sakstein:2015jca,Zumalacarregui:2010wj}, black holes \cite{Koivisto:2015mwa,Erices:2021uyu}, and neutron stars \cite{Ikeda:2021skk}. In this paper, we propose to study the stability of relativistic stars by studying  the quadratic action for both axial and polar perturbations.

The decomposition of the metric into axial and polar modes is a powerful tool for studying the resonant frequencies and damping times of gravitational waves produced by compact objects, using the quasi-normal mode (QNM) formalism \cite{kokkotas1992w,kokkotas1999quasi}. Investigation in this subject not only helps us understand the construction of neutron stars but also provides new information about modifications of general relativity (GR). However, these modifications should not suffer from ghost or gradient instabilities, as such instabilities would lead to unstable solutions to the equations that describe gravity. The stability of relativistic stars has been studied in several models belonging to scalar-tensor theories, including Horndeski theories \cite{Kase:2020yjf,Kase:2021mix}, Gauss-Bonnet couplings \cite{Minamitsuji:2022tze}, and scalar-tensor theories with a nonminimal coupling \cite{Kase:2020qvz}. 

The  paper is organized as follows. In the next section, we review the formalism of scalar-tensor theories with two metrics linked to each other via a disformal transformation, then we derive the main equations for a static and spherically symmetric configuration. We also derive the asymptotic  behavior of the metric, scalar field and  matter at the center of the NSs and at large distance. In  section \ref{2}, we extend  our analysis by considering small perturbations around the static and spherical metric. Then, after we split the perturbed metric into odd- and even-parity modes, we determine the conditions to avoid ghost and gradient instabilities. In section \ref{3}, we solve the system for two realistic equations of state, and obtain a continuum of neutron star solutions parametrized by their central energy density. We finally give some conclusions and perspectives in the final section.

\section{Neutron stars in scalar-tensor theories with disformal coupling:}\label{1}

 In this section, we will derive the background equations for scalar-tensor theories in which the physical metric $\tilde{g}_{\alpha\beta}$ is coupled to a geometric metric $g_{\alpha\beta}$ through the disformal transformation
\begin{eqnarray}\label{disformaltransformation}
\tilde{g}_{\alpha\beta}&=& \text{C}(\varphi)\;\left(g_{\alpha\beta} + \text{D}(\varphi)\partial_{\alpha}\varphi\partial_{\beta}\varphi\right),
\end{eqnarray}
where $\text{C}$ and $\text{D}$ are arbitrary functions of $\varphi$. The metric minimally coupled to matter is denoted  $\tilde{g}_{\alpha\beta}$ and is called the  Jordan frame metric. The metric on the right side of Eq.(\ref{disformaltransformation}) $g_{\alpha\beta}$, referred to as the metric in Einstein frame, is governed by an Einstein-Hilbert action.
\subsection{The model}
The total action for scalar-tensor theories with disformal coupling in Einstein frame considered here is written as
\begin{eqnarray}\label{totalaction S}
S &=& \int \, d^4 x \sqrt{-\text{g}}\left(\frac{\kappa}{2}R+\frac{\text{a}}{2}X\right)+ S_m(\tilde{g}_{\alpha\beta},\Psi),
\end{eqnarray}
where $X=\text{g}^{\alpha\beta}\partial_{\alpha}\varphi\partial_{\beta}\varphi$, $\Psi$ and $R$ are  the kinetic term, the matter fields and the Ricci scalar, respectively. $\kappa$ is a constant equal to $c^4/(8\pi G)$ with $G$  Newton's constant. We note that this theory is an extension of general relativity, which is recovered by setting $\text{C}=1$, $\text{a}=0$ and $\text{D}=0$. We obtain a subfamily corresponding to a purely conformal transformation if we set only $\text{D}=0$.

The neutron star  considered here is  described by a perfect fluid with an energy-momentum tensor $\tilde{T}_{\alpha\beta}$ of the form 
\begin{eqnarray}
\tilde{T}_{\alpha\beta}= (\tilde{\rho}+\tilde{P})\tilde{u}_\alpha\tilde{u}_\beta + \tilde{P}\tilde{g}_{\alpha\beta},
\end{eqnarray}
where $\tilde{u}^\alpha$, $\tilde{\rho}$ and $\tilde{P}$ correspond to the four-dimensional velocity vector, the energy density and the pressure of the matter in Jordan frame, respectively. The perfect fluid description is phenomenological and is usually included directly in the equations of motions. 
  However, it is often very convenient to start from an action to derive the equations of motion. In addition, by expanding the action up to second order in the perturbations around some background solution, one can determine the equations of motion for the linear perturbations of both metric and matter. Defining an action for a perfect fluid is not obvious but, fortunately, some variational formulations for a perfect fluid have been proposed in the literature \cite{Taub:1954zz,schutz1970perfect,schutz1977variational,DeFelice:2009bx,brown1993action,Bailyn:1980zz,kase2020stability}. Here we will use the Schutz-Sorkin action, given by 
\begin{eqnarray}\label{Sm}
S_m &=& \int \, d^4 x \sqrt{-\tilde{g}} \tilde{P}\left[\tilde{\mu}\right],
\end{eqnarray}
where  the chemical potential  $\tilde{\mu}$  is expressed  as
 \begin{eqnarray}
 \tilde{\mu}^2= -\tilde{g}^{\alpha\beta}\left(\partial_\alpha \tilde{q} + A\partial_\alpha B\right)\left(\partial_\beta \tilde{q} + A\partial_\beta B\right),\label{tmu2}
 \end{eqnarray}
  where  $\tilde{q}$, $A$ and $B$ are scalars fields. This relation follows from  the parametrization of the fluid four-dimensional velocity vector \cite{schutz1970perfect,brown1993action}
  \begin{eqnarray}\label{u}
\tilde{u}_\alpha &=& \frac{1}{\tilde{\mu}}\left(\partial_\alpha \tilde{q}+ A \partial_\alpha B\right).
\end{eqnarray}
 We note that the Lagrangian density $\tilde{P}\left[\tilde{\mu}\right]$ corresponds to the equation of state for a single perfect fluid. 
Here,  we considered that the entropy per particle is constant, i.e. the fluid is at  equilibrium, and thus the pressure depends only on the chemical potential. The equations of motion are obtained by varying the action (\ref{Sm}) with respect to $\tilde{g}_{\alpha\beta}$, $\tilde{q}$,  $A$ and $B$ (See Appendix \ref{App0}). In the following sections, we will explore how this formulation enables us to derive the background equations and first-order perturbed equations from both the non-perturbed and second-order action  $S$, respectively.
\subsection{Background equations}
 Now, let's suppose a static and spherical symmetric spacetime described by the Einstein frame metric
\begin{eqnarray}\label{ds0}
ds^2 =g_{\alpha\beta}dx^\alpha dx^\beta= -f(r) dt^2 + h(r) dr^2 + r^2 \left(d\theta^2 + \sin^2\theta d\phi^2\right).
\end{eqnarray}
 At this level, we also suppose that the scalar field and the thermodynamic variables depend only on the radial coordinate $r$, i.e. $\varphi\equiv\varphi(r)$, $\tilde{P}\equiv\tilde{P}(r)$, $\tilde{\mu}\equiv\tilde{\mu}(r)$....etc. The four-dimensional velocity vector in the space time (\ref{ds0}) is deduced, from the normalization $\tilde{u}_\alpha\tilde{u}_\beta \tilde{g}^{\alpha\beta}=-1$, as
\begin{eqnarray}
\tilde{u}_\alpha = \left\lbrace \sqrt{\text{C}f},0,0,0 \right\rbrace .
\end{eqnarray}
Since the background four-dimensional vector velocity is irrotational, one can choose in the background
\begin{eqnarray}
A=0,\quad B=0. 
\end{eqnarray}
Therefore, integrating the component $t$ of Eq.(\ref{u}) with respect to $t$, gives
\begin{eqnarray}
\tilde{q}(t,r)&=& -\sqrt{\text{C}}\sqrt{f}\tilde{\mu} t.
\end{eqnarray}
Then substituting this result in the component $r$ of  Eq.(\ref{u}), we obtain, from $\tilde{u}_r=0$, the following constraint
\begin{eqnarray}\label{muequation}
\frac{\tilde{\mu}'}{\tilde{\mu}}&=& -\frac{  \text{C}_\varphi \varphi'}{2 \text{C}}-\frac{f'}{2 f},
\end{eqnarray}
where the prime denotes the radial derivative and the subscript $\varphi$ represents a derivative with respect to $\varphi$. Multiplying (\ref{muequation}) by $\tilde{P}_{\tilde{\mu}}$ and using the equations $\tilde{P}'=\tilde{\mu}'\tilde{P}_{\tilde{\mu}}$ and $\tilde{\rho} = \tilde{\mu}\tilde{P}_{\tilde{\mu}}-\tilde{P}$ (where the latter equation is a result of the definitions (\ref{mutermo}) and (\ref{dtP})), we find the  matter conservation equation in the background,
\begin{eqnarray}\label{dmut}
\frac{\tilde{P}'}{\tilde{P}+\tilde{\rho}}&=&-\frac{  \text{C}_\varphi \varphi'}{2 \text{C}}-\frac{f'}{2 f}.
\end{eqnarray}
In our description, the disformal function $\text{D}$ does not appear in  the matter  conservation equation and the  term proportional to $\varphi'$  disappears if we reexpress (\ref{dmut}) in terms of $\tilde{f}= \text{C} f$.  Note that the above equation is independent of $\text{D}$ because the metric coefficients $f$, $h$ and the scalar field $\varphi$ are time-independent functions. Indeed, since the matter is conserved in Jordan frame and  $f$ equivalent to $\tilde{f}\equiv\text{C}f$, the equation written in terms of the Einstein frame metric   and the matter in Jordan frame does not depend on $\text{D}$.

We notice that the fluid dynamical variables are constrained by (\ref{muequation}). Therefore, before deriving  the equations of motion, we first substitute the metric (\ref{ds0}) in the action (\ref{totalaction S}) and then we add the constraint (\ref{muequation}) to the total action using the Lagrange multiplier $\lambda$. Doing so, we write the new action as
\begin{eqnarray}\label{s0}
S_{0}&=&\int dr \left[\frac{\sqrt{f} }{h^{3/2}}\left(\kappa( r h'+(h-1) h )-\text{a}\,h r^2 \, \varphi '^2\right)+\sqrt{f} r^2 \text{C}^{2} \tilde{P} \sqrt{h + \text{D} \varphi '^2}+ \lambda\, (\tilde{\mu}\sqrt{\text{C}}\sqrt{f})'\right].\nonumber\\
\end{eqnarray}
Varying $S_0$ with respect to $\tilde{\mu}$, we obtain
\begin{eqnarray}\label{lamda}
\lambda' &=& r^2  \text{C}^{3/2} \frac{\partial \tilde{P}}{\partial \tilde{\mu}} \sqrt{h+ \text{D} \varphi'^2}.
\end{eqnarray}
By eliminating $\lambda'$ (using the above expression) in the Euler Lagrange equations for $f$ and $h$, it follows that
\begin{eqnarray}
\frac{h'}{h}&=& -\frac{h-1}{r}+\frac{\text{a}r}{\kappa } \varphi'^2+\frac{r }{\kappa }h^{1/2} \text{C}^2 \sqrt{h+\text{D} \varphi'^2}\tilde{\rho} ,\label{eh}\\
\frac{f'}{f}&=&\frac{h-1}{r}+\frac{\text{a} r}{\kappa }\varphi'^2+\frac{r h^{3/2}  \text{C}^2}{\kappa \sqrt{h+\text{D}\varphi'^2}}\tilde{P}.\label{ef}
\end{eqnarray}
The equation of motion for the scalar field in the background is obtained by varying (\ref{s0}) with respect to $\varphi$:
\begin{eqnarray}
&&\varphi '' \left(1-\frac{h^{3/2}  \text{C}^2 \text{D}}{2 \text{a} \left(h+\text{D} \varphi '^2\right)^{3/2}}\tilde{P}\right)+h' \left(\frac{\sqrt{h}  \text{C}^2 \text{D} \varphi '}{4 \text{a} \left(h+\text{D} \varphi '^2\right)^{3/2}}\tilde{P}-\frac{\varphi '}{2 h}\right)+f' \left(\frac{\sqrt{h}  \text{C}^2 \text{D} \varphi '}{4 f \text{a} \sqrt{h+\text{D} \varphi '^2}}\tilde{\rho} +\frac{\varphi '}{2 f}\right)\nonumber\\
&&+\frac{2 \varphi '}{r}-\frac{\sqrt{h} \text{C}^2 \text{D} \varphi '}{r \text{a} \sqrt{h+\text{D} \varphi '^2}}\tilde{P} +\frac{h^{3/2}  \text{C} (3 \tilde{P}-\tilde{\rho} ) \text{C}_\varphi}{4 \text{a} \sqrt{h+\text{D} \varphi '^2}}-\frac{h^{3/2} \text{C}^2 \varphi '^2 \text{D}_\varphi}{4 \text{a} \left(h+\text{D} \varphi '^2\right)^{3/2}} \tilde{P}=0.\label{ephi}
\end{eqnarray}
Note that the radial derivatives of $f$ and $h$ can be  eliminated from the above equation by using Eqs. (\ref{eh}) and (\ref{ef}). Hence, for a particular expression of the functions $\text{C}$ and $\text{D}$, one can integrate numerically the equations (\ref{eh}), (\ref{ef}) and (\ref{ephi}).

\subsection{Asymptotic behaviors:}

 It is difficult to find an analytic solution of   Eqs (\ref{eh}), (\ref{ef}) and (\ref{ephi}), but one can  find the asymptotic solutions at the center of the star and at large distance.

\subsubsection{At large distance:}\label{larger}
Since there is no matter at large distance, the energy density and the pressure  vanish and the metric in  the two frames approach flat spacetime metric. Indeed, the asymptotic behaviors of the scalar field and the metric in Einstein frame  can be expanded as follows
\begin{eqnarray}
f =1+\sum_{i \geq 1} \frac{f_{i}^{\infty}}{r^i},\qquad h =\sum_{i \geq 0} \frac{h_{i}^{\infty}}{r^i}\quad\text{and}\quad \varphi =\sum_{i \geq 0}\frac{\varphi_{i}^{\infty}}{r^i},\label{dev1}
\end{eqnarray}
where $f_{i}^{\infty}$, $h_{i}^{\infty}$ and $\varphi_{i}^{\infty}$ are real constants (note that we have chosen $f_{0}^{\infty}=1$). Inserting these expansions in the equations of motions and then  expanding the resulting equations up to fourth order for $r\rightarrow\infty$, allows us  to determine the expression of $f_{i}^{\infty}$, $h_{i}^{\infty}$ and $\varphi_{i}^{\infty}$. Doing so, the expansions (\ref{dev1}) are explicitly given by
\begin{eqnarray}
f  &\sim & 1 +\frac{2\text{M}}{r}-\frac{\text{M}\text{Q}^2 \text{a}}{3 \kappa r^3} + \frac{2\text{M}^2\text{Q}^2 \text{a}}{3 \kappa  r^4},\\
h  &\sim & 1 -\frac{2\text{M}}{r}+\frac{\frac{5\text{M}\text{Q}^2 \text{a}}{\kappa }-8\text{M}^3}{r^3}+\frac{4\text{M}^2-\frac{\text{Q}^2 \text{a}}{\kappa }}{r^2}+\frac{16 \left(3 \kappa ^2\text{M}^4-\frac{13}{4} \kappa\text{M}^2\text{Q}^2 \text{a} +\frac{3}{16}\text{Q}^4 \text{a}^2\right)}{3 \kappa^2 r^4},\\
\varphi &\sim & \varphi_0^\infty -\frac{\text{Q}}{r}+\frac{\text{Q}\text{M}}{r^2}-\frac{\frac{4\text{M}^2\text{Q}}{3}-\frac{\text{Q}^3 \text{a}}{6 \kappa}}{r^3}+\frac{2\text{M}^3\text{Q}-\frac{2\text{M}\text{Q}^3 \text{a}}{3 \kappa }}{r^4},
\end{eqnarray}
where $\text{M}$ and $\text{Q}$ are constants of integration and they correspond to the mass of the star in Einstein frame and to the scalar field charge, respectively. In Jordan frame, the asymptotic behaviour of the function $\tilde{h}=\text{C}(h +\text{D} \varphi'^2$) is calculated as
\begin{eqnarray}
\tilde{h} &\sim & \text{C}^{\infty}-\frac{\text{Q} \text{C}^{\infty}_{\varphi}+2\text{M} \text{C}^{\infty}}{r},
\end{eqnarray}
where $\text{C}^{\infty}=\text{C}(\varphi(\infty))$. The disformal function appears at  fourth order (See Appendix \ref{app.A00}) which means that  its contribution is negligible at infinity, unlike the conformal function which appears at all orders. If we wish to have a metric in Jordan frame with $\tilde{h}\sim 1$ at large distance, we must impose a function $\text{C}^{\infty}\sim 1$ when $r\rightarrow\infty$. In this case the physical mass of the neutron star is
\begin{eqnarray}
\tilde{M}= \lim_{r\rightarrow\infty}   \frac{ r(\tilde{h}-1)}{2} =\text{M}+\frac{1}{2}\text{Q} \text{C}^{\infty}_{\varphi}.
\end{eqnarray}
In a purely disformal transformation $\text{C}=1$, we find that the masses of the star in the two frames are identical.

\subsubsection{At the center of the star:}
In order to establish  the boundary conditions at the center of the neutron star, we need to derive the behavior of $f$, $h$, $\tilde{P}$ and $\varphi$ when $r$ is close to $0$.  At the center of the star, these quantities must exhibit a regular behavior. In other words, $f$, $h$, $\tilde{P}$ and $\varphi$ must take the form

\begin{eqnarray}
f =f_{0}^{c}+\sum_{i \geq 2} f_{i}^{c}r^i,\qquad h =h_{0}^{c}+\sum_{i \geq 2} h_{i}^{c}r^i,\qquad \tilde{P} =\tilde{P}_{0}^{c}+\sum_{i \geq 2} \tilde{P}_{i}^{c}r^i\quad\text{and}\quad \varphi =\varphi_{0}^{c}+\sum_{i \geq 2} \varphi_{i}^{c}r^i.\label{dev2}
\end{eqnarray}
We impose that the first order derivative of these functions vanishes at the center. From the equations of motion, we find that these polynomials, up to  second order, are given by
\begin{eqnarray}
f &\sim & f_{0}^{c}+\frac{f_{0}^{c} (3 \tilde{P}^c_0+\tilde{\rho}^c_0)}{6 \kappa } (\text{C}^{c}r)^2,\\
h & \sim & 1+ \frac{ (\tilde{\rho}^c_0)}{3 \kappa } (\text{C}^{c}r)^2,\\
\varphi &\sim & \varphi_{0}^{c}+\frac{(\tilde{\rho}^c_0-3 \tilde{P}^c_0) (\text{C}^{c}_\varphi/\text{C}^c )}{24 \text{a}-12 \tilde{P}^c_0 (\text{C}^{c})^2 \text{D}^c} (\text{C}^{c}r)^2,\label{varphic}\\
\tilde{P} &\sim & \tilde{P}_{0}^{c}-\frac{1}{24} (\tilde{P}^c_0+\tilde{\rho}^c_0) \left(\frac{(\tilde{\rho}^c_0-3 \tilde{P}^c_0) (\text{C}^{c}_\varphi/\text{C}^{c} )^2}{2 \text{a}-\tilde{P}^c_0 (\text{C}^c)^2 \text{D}^{c}}+\frac{2  (3 \tilde{P}^c_0+\tilde{\rho}^{c}_{0})}{\kappa }\right)(\text{C}^{c}r)^2.\label{Pc}
\end{eqnarray}
In the last two approximations, the Taylor expansion breaks down for high pressure ($\lvert 1- \tilde{P}^c_0 (\text{C}^{c})^2 \text{D}^c/(2\text{a})\rvert \sim 1$) which is due to the singularity that appears for high pressure and is absent only for $\text{D}=0$ (in this case  $\text{D}^c =0$ which reduces the denominator to $2\text{a}$). However, we can overcome this problem by considering  low values of the functions $\text{D}$ such as $\lvert 1- \tilde{P}^c_0 (\text{C}^{c})^2 \text{D}^c/(2\text{a})\rvert \ll 1 $. Note that the same singularity appears in the Jordan frame for $\tilde{h}$,
\begin{eqnarray}
\tilde{h} &\sim & \text{C}^c+\left(\frac{4 \text{C}^{c} \text{D}^{c} (\tilde{\rho}^{c}_0-3 \tilde{P}^{c}_0)^2 (\text{C}^{c}_\varphi)^2}{\left(24 \text{a}-12 \tilde{P}^{c}_0 (\text{C}^{c})^2 \text{D}^{c}\right)^2}+\frac{ (\tilde{\rho}^{c}_0-3 \tilde{P}^{c}_0) (\text{C}^{c}_\varphi/\text{C}^{c})^2}{24 \text{a}-12 \tilde{P}^{c}_0 (\text{C}^{c})^2 \text{D}^{c}}+\frac{\tilde{\rho}^{c}_0(\text{C}^{c})}{3 \kappa }\right)( \text{C}^c r)^2.
\end{eqnarray}
In the case of a purely disformal transformation $\text{C}=1$, this approximation is reduced to
\begin{eqnarray}
\tilde{h} &\sim &1+\frac{\tilde{\rho}^{c}_0}{3 \kappa }  r^2,
\end{eqnarray}
which corresponds to  GR, and thus the function $\text{D}$ does not affect the behavior of the metric at the center of the star. The above approximations (at large and small distances) for  matter and  metric in both frames, were also found in Ref.\cite{minamitsuji2016relativistic}.
\section{Ghost and gradient instabilities:}\label{2}
 To investigate the ghost and gradient instabilities of linear perturbations, we consider the perturbed metric
\begin{eqnarray}
ds_{tot}^2 &=& ds^2 + ds_{per}^2,
\end{eqnarray}
where $ds_{per}^2$ can be  decomposed into  polar and axial parts. The polar part has even parity while the axial one has odd parity when performing a rotation in the two dimensional subspace $(\theta,\phi)$. This decomposition will allow us to split  the  first order perturbation equations   into axial and polar equations. In this paper, for  even perturbations, we will  choose the uniform curvature gauge \cite{kase2020stability}
\begin{eqnarray}
ds_{per,even}^2 &= f H_0^{lm} Y_{lm}dt^2+2\sqrt{fh} H_1^{lm} Y_{lm}dtdr+h H_2^{lm} Y_{lm}dr^2\nonumber\\
&+2\sqrt{h}H_5^{lm} (\partial_\theta Y_{lm} r d\theta +\partial_\phi Y_{lm} r\sin\theta d\phi)dr,
\end{eqnarray}
where $H_0^{lm}$, $H_1^{lm}$, $H_2^{lm}$ and $H_5^{lm}$ depend on the coordinates $t$ and $r$, and $Y_l^m(\theta,\phi)$ are the spherical harmonic functions. For  axial perturbations, we choose  the Regge-Wheeler gauge \cite{kase2020stability}
\begin{eqnarray}
ds_{per,odd}^2 &=& 2 r^2\sin \theta \left(\partial_\theta Y_{lm}+\frac{1}{\sin\theta}\partial_\phi Y_{lm}\right) \left(h_1^{lm} dtd\theta +h_0^{lm} dtd\phi \right),
\end{eqnarray}
where $h_0^{lm}$ and $h_1^{lm}$ are functions of the coordinates $t$ and $r$. For polar perturbations, the scalar field  and $\tilde{q}$ are perturbed, and read, in terms of the spherical harmonics $Y_{lm}$, as
\begin{eqnarray}\label{scalarper}
\varphi = \varphi(r)+Y_{lm}\,\delta\varphi^{lm},\;\;\text{and}\;\; \tilde{q}= -\sqrt{\text{C}}\sqrt{f}\tilde{\mu} t+ Y_{lm}\,\delta\tilde{q}^{lm}.
\end{eqnarray}
But, they  vanish in the axial case. Since we will show in the next subsections that the scalar fields $A$ and $B$ do not contribute to the equations of motion, it is not necessary to decompose them in terms of  spherical harmonics, i.e.
\begin{eqnarray}\label{ABper}
A= \delta A(t,r,\theta,\phi)\;\;\text{and}\;\; B= \delta B(t,r,\theta,\phi).
\end{eqnarray}
 In addition, the perturbed four-dimensional vector velocity's components are decomposed as
\begin{eqnarray}\label{muper}
\delta\tilde{u}^\alpha =\left( \begin{matrix}\delta\tilde{u}^{t,lm}Y_{lm}\\
                                       \delta\tilde{u}^{r,lm}Y_{lm}\\
                                      \frac{\sqrt{f}}{\text{C}r^2} v_p \partial_\theta Y_{lm}+v_a	\frac{1}{\sin\theta}\partial_\phi Y_{lm}\\
                                        v_a\partial_\theta Y_{lm}+\frac{\sqrt{f}}{\text{C}r^2}v_p \frac{1}{\sin\theta}\partial_\phi Y_{lm}

             \end{matrix}\right)+\delta^2 \tilde{u}^\alpha(t,r,\theta ,\phi),
\end{eqnarray}
where $v_a$ and  $v_p$, which are functions of $t$ and $r$,  represent the axial and polar angular velocities of the fluid, respectively.   The second order perturbation is essential to satisfy the condition
\begin{eqnarray}\label{noramalisation2}
\delta^2 \left[\tilde{u}_\alpha\tilde{u}_\beta \tilde{g}^{\alpha\beta}\right]=0.
\end{eqnarray}
 Finally, using the Eqs. (\ref{u}) and (\ref{noramalisation2}), we deduce 
 \begin{eqnarray}\label{mu2}
 \tilde{\mu} &= \tilde{\mu}(r)\left(1+\delta\tilde{\mu}^{lm}Y_{lm}+\delta^2 \tilde{\mu}(t,r,\theta,\phi)\right).
 \end{eqnarray}
 
 We will give the expression of $\delta^2 \tilde{\mu}$  for the polar and the axial cases in the next subsections. In the following, since the integer $m$ will  not contribute to the perturbed equations, we will set $m=0$ and  we remove the (sub/super)script "$lm$" from the perturbed quantities.
\subsection{Polar perturbations:}

 For  the even parity sector, we expand the definition (\ref{u}) using  the expansions (\ref{scalarper}), (\ref{muper}) and (\ref{mu2}) up to  second order. At  first order,  the components $r$, $t$ and $\theta$ give the equations
\begin{eqnarray}
\frac{\delta\tilde{q}'}{\tilde{\mu}\text{C}^{1/2}}&=& \frac{\text{D}   \varphi '}{\sqrt{f}}\delta\dot{\varphi}+\sqrt{\text{C}} \delta\tilde{u}_r \left(\text{D} \varphi '^2+h\right)+\sqrt{h} H_1, \label{c1}\\
\frac{\delta \dot{\tilde{q}}}{\tilde{\mu}\sqrt{f}\text{C}^{1/2}}&=& -\frac{\text{C}_{\varphi }}{2 \text{C}}\delta \varphi  -\delta\tilde{\mu}+\frac{H_0}{2}  ,\label{c2}\\
\frac{l(l+1)}{r^2 \tilde{\mu} }\frac{\delta\tilde{q}}{\text{C}^{1/2}} &=&  l(l+1)\,v_p,\label{c3}
\end{eqnarray}
where the dot stands for the time derivative. At  second order, from the component $t$ of Eq.(\ref{u}), we derive the equation
\begin{eqnarray}
\delta^2 \tilde{\mu} &=& \frac{1}{2} Y_{l0}^2 \left(-\frac{1}{2} \text{C}  \left(h+\text{D} \varphi '^2\right)\delta\tilde{u}_r^2+\frac{ \text{C}_{\varphi }^2}{8 \text{C}^2}\delta\varphi^2+\delta\tilde{\mu} \left(\frac{1}{2} H_0-\frac{  \text{C}_{\varphi }}{2\text{C}}\delta\varphi\right)-\frac{\text{C}_{\varphi }}{4\text{C}}  \delta \varphi  H_0+\frac{1 }{8}H_0^2\right)\nonumber\\
&&-\frac{f}{2r^2} \partial_\theta Y_{l0}^2 v_p^2-\frac{1 }{\sqrt{f}\sqrt{\text{C}} \tilde{\mu}}\delta A \delta\dot{B}.\label{mut2p}
\end{eqnarray}
However, from the equations (\ref{cnstrntaB}), we have 
\begin{eqnarray}
\delta\dot{A}=0\;\;\text{and}\;\;\delta\dot{B}=0\;\;\Leftrightarrow\;\;\delta A=\delta A(r,\theta,\phi)\;\;\text{and}\;\;\delta B= \delta B(r,\theta,\phi).
\end{eqnarray}
We can deduce that the expansion (\ref{mu2}) is independent of the functions $A$ and $B$. Thus, they will not  appear in the equations of motion. Combining Eq. (\ref{c1}) with Eq. (\ref{c3}), Eq. (\ref{c2}) with Eq. (\ref{c3}) and Eq. (\ref{c1}) with Eq. (\ref{c2}), gives, respectively, the following equations
\begin{eqnarray}
&& E_0\equiv v_p'-\frac{\sqrt{\text{C}}}{\sqrt{f}} \left(h+\text{D} \varphi '^2\right)\delta\tilde{u}^r -\frac{\text{D} \varphi ' \delta\dot{ \varphi}}{f}-\frac{\sqrt{h}}{\sqrt{f}} H_1 =0,\label{c4}\\
&& E_1\equiv \dot{v}_p+\delta\tilde{\mu}+\frac{\text{C}_\varphi}{2\sqrt{\text{C}}} \delta\varphi  -\frac{1}{2} H_0  =0,\label{c5}\\
&& E_2\equiv (\delta\tilde{\mu}+\frac{\text{C}_\varphi}{2\text{C}}\delta\varphi)'+\frac{\sqrt{\text{C}}  \left(h+\text{D} \varphi '^2\right)}{\sqrt{f}}\dot{\delta\tilde{u}}^r+\frac{\text{D} \varphi '}{f}\ddot{ \delta \varphi}+\frac{\sqrt{h} }{\sqrt{f}}\dot{H}_1-\frac{1}{2} H_0'=0.\label{c6}
\end{eqnarray}
We note that the equations above are not independent from each other, since one can derive $E_2$ by subtracting the radial derivatives of $E_1$ and the time derivative of $E_0$ to eliminate $v_p$. Moreover, the equations $E_1$ and $E_2$ can be also derived from the energy momentum tensor conservation equation, where its  $r$ and $\theta$ components are equivalent to  $E_2$ and $E_1$, respectively.   We choose to include  Eq. (\ref{c6}) and (\ref{c5}) as constraints in the second order expansion of the matter action ($\delta^2 S_m$) using two Lagrange multipliers $\delta\lambda_1$ and $\delta\lambda_2$. Doing so, the second-order matter action is written as
\begin{eqnarray}
\delta^{2}S_{m} &=&\int drdt\left(\delta^2\sqrt{-\tilde{g}} \tilde{P}+\delta\sqrt{-\tilde{g}} \delta\tilde{P}+\sqrt{-\tilde{g}} \delta^2\tilde{P}\right)-\delta\lambda_2 E_2-l(l+1)\delta\lambda_1 E_1.
\end{eqnarray}
If we vary $\delta^{2}S_{m}$ with respect to the functions $\delta\tilde{\mu}$, $v_p$ and $\delta\tilde{u}^r$, the same functions  can be written, by solving the resulting equations, as
\begin{eqnarray}
\delta\tilde{\mu} &=& -\frac{\tilde{c}_m^2 h}{2(h+\text{D}\varphi'^2)}H_2-\frac{\tilde{c}_m^2\tilde{\mu}}{r^2\text{C}^{3/2}(\tilde{P}+\tilde{\rho})\sqrt{h+\text{D}\varphi'^2}}(l(l+1)\delta\lambda_1/\sqrt{f}+ (\delta\lambda_2/\sqrt{f})')-\frac{\text{D}\tilde{c}_m^2 \varphi'}{h+\text{D}\varphi'^2}\delta\varphi'\nonumber\\
&&+\left(\frac{\text{C}_\varphi}{\text{C}}\left(1-\frac{3\tilde{c}_m^2}{2}\right)-\frac{\tilde{c}_m^2\text{D}_\varphi \varphi'^2}{2(h+\text{D}\varphi'^2)}\right)\delta\varphi,\\
\delta\tilde{u}^r &=& \frac{\tilde{\mu}}{r^2\text{C}^{3/2}f(\tilde{P}+\tilde{\rho})\sqrt{h+\text{D}\varphi'^2}}\dot{\delta\lambda}_2,\\
v_p               &=& -\frac{\tilde{\mu}}{\text{C}^{3/2}f^{3/2}(\tilde{P}+\tilde{\rho})\sqrt{h+\text{D}\varphi'^2}}\dot{\delta\lambda}_1.
\end{eqnarray}

 Now, if we substitute these expressions into the action ($\delta^{2}S_{m}$), it will depend only on the metric and the Lagrange multipliers. By considering the second order expansion of the total action (\ref{totalaction S}) and after integrating by parts, it follows that

\begin{eqnarray}\label{S2polar}
\delta^{2}S^{polar} &=&\int drdt\left(H_0\left(a_1 \delta\varphi' +L a_2 H_5' +L a_3 H_2'+a_4 \delta\varphi +a_5  H_5 +a_6 H_2+ f_1 (\delta\tilde{\lambda}_2'+L \frac{1}{\sqrt{f}}\delta\lambda_1)\right)\right.\nonumber\\
 &&\left. +L a_7 H_1^2+H_1(f_2 \dot{\delta\tilde{\lambda}}_2+a_8\delta\dot{\varphi} +L a_{9} \dot{H}_5 + a_{10} \dot{H}_2)+ a_{11} H_2^2+H_2(a_{12}\delta\varphi' +a_{13}\delta\varphi  \right.\nonumber\\ 
 &&\left. +L a_{14}H_5 +f_3 (\delta\tilde{\lambda}_2'+L \frac{1}{\sqrt{f}}\delta\lambda_1))+ L a_{15}H_5^2 + L a_{16}\dot{H}_5^2 + L a_{17}H_5 \delta\varphi + e_1 \delta\varphi^2 + e_2 \delta\varphi'^2\right.\nonumber\\ 
 &&\left.  + e_3 \delta\dot{\varphi}^2+e_4(\delta\tilde{\lambda}_2'+L \frac{1}{\sqrt{f}}\delta\lambda_1)\delta\varphi +L c_1 \dot{\delta\lambda}_1^2+c_2 \dot{\delta\tilde{\lambda}}_2^2 +c_3 \delta\tilde{\lambda}_2'^2 +L c_4 \delta\lambda_1 \delta\tilde{\lambda}_2'+L^2 c_5 \delta\lambda_1^2 \right),\nonumber\\ 
\end{eqnarray}
with
\begin{eqnarray}
\delta\tilde{\lambda}_2 &=&\frac{1}{f^{1/2}} \delta\lambda_2 + \frac{\text{D}\text{C}^{3/2}(\tilde{P}+\tilde{\rho})\varphi' r^2}{\tilde{\mu}\sqrt{h+\text{D}\varphi'^2}}\delta\varphi ,
\end{eqnarray}
where $L=l(l+1)$ and the coefficients $a_i$, $c_i$, $f_i$ and $e_i$ are given in  appendix  \ref{app.A}. 
\subsubsection{The case $l\geq 2$:}\label{lsup2}
In order to rewrite the action (\ref{S2polar}) in the form of  a wave action, we need to eliminate the non-dynamical variables. To do so, we vary (\ref{S2polar}) with respect to $H_0$ and $H_1$, respectively, to obtain
\begin{eqnarray}
&& a_1 \delta\varphi' +L a_2 H_5' + a_3 H_2'+a_4 \delta\varphi +a_5  H_5 +a_6 H_2+ f_1 (\delta\tilde{\lambda}_2'+L \frac{1}{\sqrt{f}}\delta\lambda_1)=0,\\
&& 2 L a_7 H_1+f_2 \dot{\delta\tilde{\lambda}}_2+a_8\delta\dot{\varphi} +L a_{9} \dot{H}_5 + a_{10} \dot{H}_2=0,
\end{eqnarray}
and we define  the combination, which will allow us to express the dynamics of the gravitational sector \cite{DeFelice:2011ka,Kobayashi:2014wsa,kase2020stability,minamitsuji2016relativistic}, as
\begin{eqnarray}
&& \psi = L a_2 H_5 + a_3 H_2.
\end{eqnarray}
The last three  equations are solved with respect to $H_1$, $H_2$ and $H_5$. Then, substituting the obtained solutions  in (\ref{S2polar}), the time derivative of $H_5$ and $H_2$ are eliminated from the action. Therefore, the only functions left in (\ref{S2polar}) are $\delta\lambda_1$, $\delta\tilde{\lambda}_2$, $\psi$ and $\delta \varphi$. After long calculations, we arrive to
\begin{eqnarray}
\delta^{2}S^{polar} &=& \int drdt\;\left( \dot{\chi}^t\textbf{K}\dot{\chi}+{\chi^{t}}' \textbf{G}\chi' +\chi^t \textbf{L}\chi' +\chi^t \textbf{M}\chi\right),
\end{eqnarray} 
with
\begin{eqnarray}
{\chi}^t=\{\delta\lambda_1,\delta\tilde{\lambda}_2,\psi,\delta \varphi\},
\end{eqnarray}
and $\textbf{K}$, $\textbf{G}$ and $\textbf{M}$ are $4\times 4$  matrices where $\textbf{G}_{11}=0$ and $\textbf{M}_{22}=0$. The other components have a complicated expression, but due to the background equations the ghost-free conditions are simplified. The ghost instability is absent if the matrix $\textbf{K}$ is  positive definite, i.e.
\begin{eqnarray}
\textbf{K}_{11}&=& \frac{L\text{C}\tilde{\mu}^2 }{2 f^{3/2}(P+\rho)(h+\text{D}\varphi'^2)} \geq 0,\label{L1}\\
\sum_{\{i,j\}=\{1,2\}}\epsilon^{ij}\textbf{K}_{1i}\textbf{K}_{2j}&=& f^2 \tilde{\mu}^4\text{C}^2\left(L r^4 \left(\text{a}^2 \varphi '^4+\varphi'^2 \left(2 \text{a} h^{3/2} \text{P}+\text{D} h^2 \rho  (\text{P}+\rho )\right)+h^3 \rho ^2\right)\right.\nonumber\\
&& \left. +2 \kappa  r^2 \left(\text{a} L (h L+h-3) \varphi'^2+h^{3/2} ((L-2) \text{P}-\rho  (L (h L+h-4)+2))\right)\right.\nonumber\\
& &\left.
+\kappa ^2 L (h L+h-3)^2\right)/\Delta\geq 0,\label{L2}\\
\sum_{\{i,j,n\}=\{1,2,3\}}\epsilon^{ijn}\textbf{K}_{1i}\textbf{K}_{2j}\textbf{K}_{3n}&=& 2L  f^{1/2}h^{3/2}\text{C}^{2}\tilde{\mu}^4\left(2 \kappa  (L-2)+L r^2 \varphi '^2 \left(2 \text{a}+\text{D} \sqrt{h} \rho \right)\right)/\Delta\geq 0,\label{L3}
\end{eqnarray}
where 
\begin{eqnarray}
\Delta &=& 4 f^4 L (\text{P}+\rho )^2 \left(h+\text{D} \varphi '^2\right) \left(\kappa  (h L+h-3)+\text{a} r^2 \varphi '^2+h^{3/2} \text{P} r^2\right)^2,
\end{eqnarray}
and $\epsilon^{ijk..}$ is the Levi-Civita symbol.  Since, $\text{P}+\rho\geq 0$, the first condition is automatically verified. The second one, which has a complicated expression,  is verified when the numerator is positive (since $\Delta\geq 0$).  The final condition is satisfied for
\begin{eqnarray}
2 \kappa  (L-2)+L r^2 \varphi '^2 \left(2 \text{a}+\text{D} \sqrt{h} \rho \right)\geq 0.
\end{eqnarray}
The fourth condition, to have $\textbf{K}$ positive definite, is   
\begin{eqnarray}
Det[\textbf{K}]&=& h \tilde{\mu }^4 \left(4 \text{a} \text{C}^2 h \kappa  (L-2) L-2 \text{C}^2 \text{D} \sqrt{h} L \rho  \varphi '^2 \left(\text{D} \left(\text{a} L r^2 \varphi '^2+\kappa  (L-2)\right)+\text{a} h L r^2\right)\right.\nonumber\\
& &\left. -4 \text{C} \sqrt{h} L \rho  r^3 \text{C}_{\varphi } \varphi ' \left(4 \text{a}+\text{D} \sqrt{h} \rho \right) \left(h+\text{D} \varphi '^2\right)-\text{C}^2 \text{D}^2 h L^2 \rho ^2 r^2 \varphi '^2 \left(h+\text{D} \varphi '^2\right)\right.\nonumber\\
& &\left. -4 \rho ^2 r^4 \text{C}_{\varphi }^2 \left(h+\text{D} \varphi '^2\right)^2\right)/\Delta\geq 0,
\end{eqnarray}
which is verified when the numerator is positive. 

Another feature which should be studied is  gradient instabilities. In order to have a good theory of gravity, the propagation speed squared of the vector $\chi$ must be positive in both radial and angular directions. In fact, to derive the conditions of gradient instabilities, we consider the solution $\chi=\chi_0 e^{I(\omega t- k r- l\theta)}$, where $\chi_0$ is a constant vector, and $\omega$ and $k$ are the frequency and wavenumber, respectively. If we wish to check the absence of  gradient instability in the radial direction or in the angular direction, we take the limits $\omega\rightarrow\infty$ and $k\rightarrow\infty$ or we take the limits $\omega\rightarrow\infty$ and $l\rightarrow\infty$, respectively. Then, to ensure non-vanishing solutions, we impose 
\begin{eqnarray}
Det[f\omega^2\textbf{K}+h k^2\textbf{G}]=0,
\end{eqnarray}
for the radial direction and  
\begin{eqnarray}
Det[L f\omega^2\textbf{K}+r^2\textbf{M}]=0,
\end{eqnarray}
for the angular directions. The interesting result of our calculation is that we find that the radial propagation speed ($c_{r_2}=\omega/k$) of $\delta\lambda_1$ and the angular propagation speed ($c_{\Omega_1}=\omega/l$) of $\delta\tilde{\lambda}_2$ vanish, but the radial propagation speed of $\delta\tilde{\lambda}_2$ and the angular propagation speed of $\delta\lambda_1$ are given by
\begin{eqnarray}
c_{r_2}^2=\frac{h}{h+\text{D}\varphi'^2}\tilde{c}_m^2,\qquad c_{\Omega_1}^2=\tilde{c}_m^2.
\end{eqnarray}
The other solutions describe the propagating speed of $\delta\varphi$ and $\psi$, for which  the gradient instabilities are avoided if
\begin{eqnarray}
c_{r_\pm}^2 &=&\frac{A_2}{2A_1}\left(1\pm\sqrt{1-\frac{A_3A_1}{4A_2^2}}\right)\geq 0,\\
c_{\Omega\pm}^2 &=&\frac{B_2}{2B_1}\left(1\pm\sqrt{1-\frac{B_3B_1}{4B_2^2}}\right)\geq 0,
\end{eqnarray}
where the expressions of $A_i$ and $B_i$, with $i=\{1,2,3\}$ are given in  Appendix (\ref{app.B}). Like in Horndeski theories, the propagating speeds of $\delta\varphi$ and $\psi$ in angular and radial directions are affected by the scalar field which depend on the form of the functions $\text{C}$ and $\text{D}$. Despite the complexity of the velocities $c_{r_\pm}^2$ and $c_{\Omega\pm}^2$, one can estimate their behaviors and signs at the center of the star and at the exterior of the star. In fact, the speeds in both directions are reduced to the speed of light outside the star and we calculate $c_{r_\pm}^2$ when $r$ tends to zero as
\begin{eqnarray}
c_{r_\pm}^2 = 1+O(r^2),\quad c_{\Omega\pm}^2 = 1+O(r^2),
\end{eqnarray}
which means that  gradient instabilities in  both directions are absent at the center of the star for any  functions $\text{D}$ and $\text{C}$.

\subsubsection{The case $l=0$:}

 If we impose  $l=0$ in the action (\ref{S2polar}), we must reduce the degrees of freedom by choosing the gauge $H_0=0$ (or $H_1=0$). Therefore, the action (\ref{S2polar}) is reduced to
\begin{eqnarray}
\delta^{2}S^{polar} &=& \int drdt \left(H_1(f_2 \dot{\delta\tilde{\lambda}}_2+a_8\delta\dot{\varphi} + a_{10} \dot{H}_2)+ a_{11} H_2^2+H_2(a_{12}\delta\varphi' +a_{13}\delta\varphi +f_3 \delta\tilde{\lambda}_2') +  e_1 \delta\varphi^2 \right.\nonumber\\ 
 &&\left.  + e_2 \delta\varphi'^2 + e_3 \delta\dot{\varphi}^2+e_4 \delta\tilde{\lambda}_2'\delta\varphi +c_2 \dot{\delta\tilde{\lambda}}_2^2 +c_3 \delta\tilde{\lambda}_2'^2  \right). 
\end{eqnarray}
Solving the  equations obtained from the variation of this action with respect to $H_1$ and  $H_2$ and then  replacing the result in the  action, enable us to rewrite it as 
\begin{eqnarray}
\delta^{2}S^{polar} &=& \int drdt\;\left( \dot{\chi}^t\textbf{K}\dot{\chi}+{\chi^t}' \textbf{G}\chi'+  \chi^t \textbf{L}\chi' + {\chi}^t \textbf{M}\chi\right),
\end{eqnarray} 
with
\begin{eqnarray}
{\chi}^t=\{\delta\tilde{\lambda}_2, \delta \varphi\},
\end{eqnarray}
where the matrices $\textbf{K}$ and $\textbf{G}$ are diagonal $2\times 2$ matrices. Thus, the absence of ghost is ensured by the conditions 
\begin{eqnarray}
\textbf{K}_{11}&=& \frac{\tilde{\mu}^2 \text{C}}{2  \sqrt{f} r^2 (\text{P}+\rho )}\geq 0,\\
\textbf{K}_{22}&=& \frac{\text{a} \sqrt{h} r^2}{\sqrt{f}}-\frac{ \text{D}^2 \rho  r^2 \varphi '^2}{2 \sqrt{f} }\geq 0.
\end{eqnarray}
We must impose $\text{a}\geq 0$ in order to ensure the last condition outside the star. The gradient instability condition is given by
\begin{eqnarray}
c_{r_1}^2 &=&\frac{ h}{h+\text{D} \varphi '^2}\tilde{c}_m^2\geq 0,\\
c_{r}^2 &=& \frac{\sqrt{h} \left(2 \text{a} h+\text{D} \varphi '^2 \left(2 \text{a}+\text{D} \sqrt{h} \text{P}\right)\right)}{\left(\text{D} \varphi '^2+h\right) \left(2 \text{a} \sqrt{h}-\text{D}^2 \rho  \varphi '^2\right)}\geq 0.
\end{eqnarray}

 We note that, in the case $\text{D}=0$,  the  propagation speed of the scalar field is equal to the speed of light in  vacuum. The same result is recovered at the exterior of the star. At the center of the star, the propagating speed of the scalar field behaves as
\begin{eqnarray}
c_{r}^2&=& 1+O(r^2).
\end{eqnarray}
As expected, we do not find gradient instabilities for all forms of the functions $\text{D}$ and $\text{C}$. We observe that $\text{D}$ plays a crucial role in modifying the propagation speed of the scalar field with respect to the speed of light. 

\subsubsection{The case $l=1$:}
We have seen that in the case $l\geq 0$ the propagating speed of the vector $\chi$ is not defined when $l=1$. This is due to the presence of an extra  gauge degree of freedom \cite{kase2020stability}. In this paper, we fix the gauge by setting $\delta\varphi=0$. Following the same steps as in the subsection \ref{lsup2}, we obtain the conditions for the absence of the gradient instability  
\begin{eqnarray}
c_{r_1}^2&=& 0,\\
c_{r_2}^2&=& \frac{ h}{h+\text{D} \varphi '^2}\tilde{c}_m^2\geq 0,\\
c_{r}^2&=& \frac{2 \text{a} \text{D} \varphi '^2+2 \text{a} h-\text{D} h^{3/2} \text{P}}{\left(2 \text{a}+\text{D} \sqrt{h} \rho \right) \left(h+\text{D} \varphi '^2\right)}\geq 0,
\end{eqnarray}
where the propagation speed of $\psi$  reduce to the speed of light if $\text{P}=0$ and  $\rho=0$ or if $\text{D}=0$. At $r=0$, the value of $c_{r}^2$ is 
\begin{eqnarray}\label{cr3l1}
c_{r}^2&=&\frac{2\text{a}-\tilde{P}^c_0 (\text{C}^c)^2 \text{D}^c}{2\text{a}+\tilde{\rho}^c_0  (\text{C}^c)^2 \text{D}^c}.
\end{eqnarray}
Since the pressure and the energy density of the star are positive, a gradient instability at the center of the stars occurs if
\begin{eqnarray}
\tilde{P}^c_0 (\text{C}^c)^2 \text{D}^c\geq 2\text{a},\quad\text{and}\quad \tilde{\rho}^c_0 (\text{C}^c)^2 \text{D}^c\geq -2\text{a}.
\end{eqnarray}
These conditions must be taken into consideration in the numerical analysis of the background equations. Finally, the no ghost conditions in this case are recovered by taking the limit $L\rightarrow 1$ of the equations (\ref{L1}), (\ref{L2}) and (\ref{L3}).


\subsection{Axial perturbations:}
For axial perturbations, we  follow the same steps as for polar perturbations, but the calculations are now simpler. From Eqs.(\ref{u}), (\ref{muper}) and (\ref{mu2}), we obtain  the constraints
\begin{eqnarray}
v_a &=& \frac{1}{\sqrt{f} \sqrt{\text{C}}}h_0,\\
\delta^2 \tilde{\mu}&=& -\frac{r^2\text{C}}{2 f } \partial_\theta Y_{l0}^2 h_0^2.\end{eqnarray}
Inserting the last equation in the total action, then  perturbing it up to second order and integrating  by parts, it follows that
\begin{eqnarray}\label{s2axial}
\delta^{2}S^{axial} &=& \int drdt \frac{\kappa  l(l+1) r^2 }{4 \sqrt{f} \sqrt{h}}\left(h (l(l+1) -2) h_0^2-f (l(l+1) -2) h_1^2+r^2 \left(h_0'-\dot{h}_1\right)^2\right).
\end{eqnarray}
As  found in Horndeski theories \cite{kase2020stability}, the fluid has no effect on the Lagrangian and we distinguish the two cases $l=1$ and $l\geq 2$.
\subsubsection{The case $l=1$:}
In this case, the action is reduced to 
\begin{eqnarray}
\delta^{2}S^{axial} &=& \int drdt \frac{\kappa  l(l+1) r^4 }{4 \sqrt{f} \sqrt{h}} \left(h_0'-\dot{h}_1\right)^2,
\end{eqnarray}
which  gives us, by using the Euler-Lagrange equations, the following  equations
\begin{eqnarray}
\ddot{h}_1-\dot{h}_0'=0,\\
\left( \frac{r^4 }{\sqrt{f} \sqrt{h}}(\dot{h}_1-h_0')\right)'=0.
\end{eqnarray}
If we fix the gauge $h_1=0$ and integrate the above equations we find
\begin{eqnarray}\label{h0-l=0}
h_0 \propto \int d\hat{r} \frac{\sqrt{f} \sqrt{h}}{\hat{r}^4}.
\end{eqnarray}
Note that we have eliminated the arbitrary function (it depends only on time) that arises in our integration with respect to $r$, by using a specific choice of gauge mode that appears in our case \cite{kase2020stability}. We observe in the result (\ref{h0-l=0}) that the axial metric perturbation is time independent which is similar to the one we find in Horndeski theories. However, this result is modified in Jordan frame as
\begin{eqnarray}
\tilde{h}_0 &\propto & \text{C}\int d\hat{r} \frac{\text{C}\sqrt{\tilde{f}} \sqrt{\tilde{h}+\text{D}\varphi '^2}}{\hat{r}^4},
\end{eqnarray}
which means that the  moment of inertia of a relativistic star is also modified in these theories, and thus one expects to  have a deviation of the relation between the mass and the moment of inertia with respect to GR (see Ref.\cite{minamitsuji2016relativistic}). 
\subsubsection{The case $l\geq 2$:}
In this case, we use the Lagrange multiplier method which allows us to have the explicit form of $h_0$ and $h_1$  in terms of  $\xi=(\dot{h}_1-h_0')/\sqrt{h}\sqrt{f}$ \cite{minamitsuji2016relativistic}. Doing so, the action (\ref{s2axial}) becomes
\begin{eqnarray}
\delta^{2}S^{axial} &= \int drdt\frac{\sqrt{f}\kappa  l(l+1) r^6  }{4 (l(l+1)-2)\sqrt{h}}\left(\left(\frac{2 r^2 \text{C}^2 \left(h (\tilde{P}-\tilde{\rho} )-\tilde{\rho}  \text{D}\varphi'^2\right)}{\kappa\sqrt{h+\text{D}\varphi'^2}   }+\frac{(6-l(l+1)) \sqrt{h}}{r^2 }\right)\xi ^2+\frac{h   }{ f }\dot{\xi}^2-\xi '^2\right).
\end{eqnarray}
Similarly to  GR,  ghost and gradient instabilities, in the radial direction are absent in the axial mode, as long as $h/f$ is positive. Also, the above action shows that the radial propagating speed of the axial modes in Einstein frame is equal to the speed of light. In the angular direction, the gradient instability is also avoided, since as $l$ tends to infinity we have
\begin{eqnarray}
c_\Omega^2 &=& 1.
\end{eqnarray}
Therefore, we have demonstrated that the ghost and gradient instabilities of the axial modes are absent in disformal scalar-tensor theories.

\section{Numerical analysis:}\label{3}

In order to confirm our analytic studies and to see if our model is stable for all values of $r$, we must solve numerically the background equations. To do so, we set $\text{a}=1$ and we consider a model symmetric under scalar field reflection (invariant under the transformation $\varphi\rightarrow-\varphi$)  with  the following functions $\text{C}$ and $\text{D}$ 
\begin{eqnarray}\label{PC}
\text{C}=e^{p\varphi^2}\quad \text{and}\quad \text{D}=\Lambda,
\end{eqnarray}
where $\Lambda$ and $p$ are constants.  In our numerical integration, we  use the dimensionless variables $x=\ln(r/r_0)$ and $\tilde{\varphi}=\varphi\sqrt{G}/c^2$ and the dimensionless constants $\tilde{p}=p c^4/G$ and $\tilde{\Lambda}=\Lambda c^2/\rho_0$, where

\begin{eqnarray}
r_0=\frac{c}{\sqrt{G \rho_0}}=89.664\, {\rm km}\,, \qquad \rho_0=m_{\rm n} n_0=1.6749\times 10^{14} {\rm g.cm}^{-3}\,,
 \end{eqnarray}
 where $m_{\rm n}$ is the neutron mass and $n_0=0.1\; {\rm fm}^{-3}$ is the typical number density  in neutron stars. Given an equation of state and a  particular  choice of the parameters $\tilde{p}$ and $\tilde{\Lambda}$,  the radial integration of the background equations depends on the central energy density $\tilde{\rho}_c$ and the central value $\varphi_0^c$. The other quantities at $r=0$ can be expressed in terms of $\rho_c$, as discussed in the second section. In addition, we impose that at $r\rightarrow \infty$ the metric coefficient $\tilde{h}$ tends to $1$, which can be satisfied only if $\varphi^\infty=0$. The latter happens only for a particular value of $\varphi_0^c$. 
 \begin{figure}[htb]
\centering
\includegraphics[width=0.4\textwidth, height=5cm]{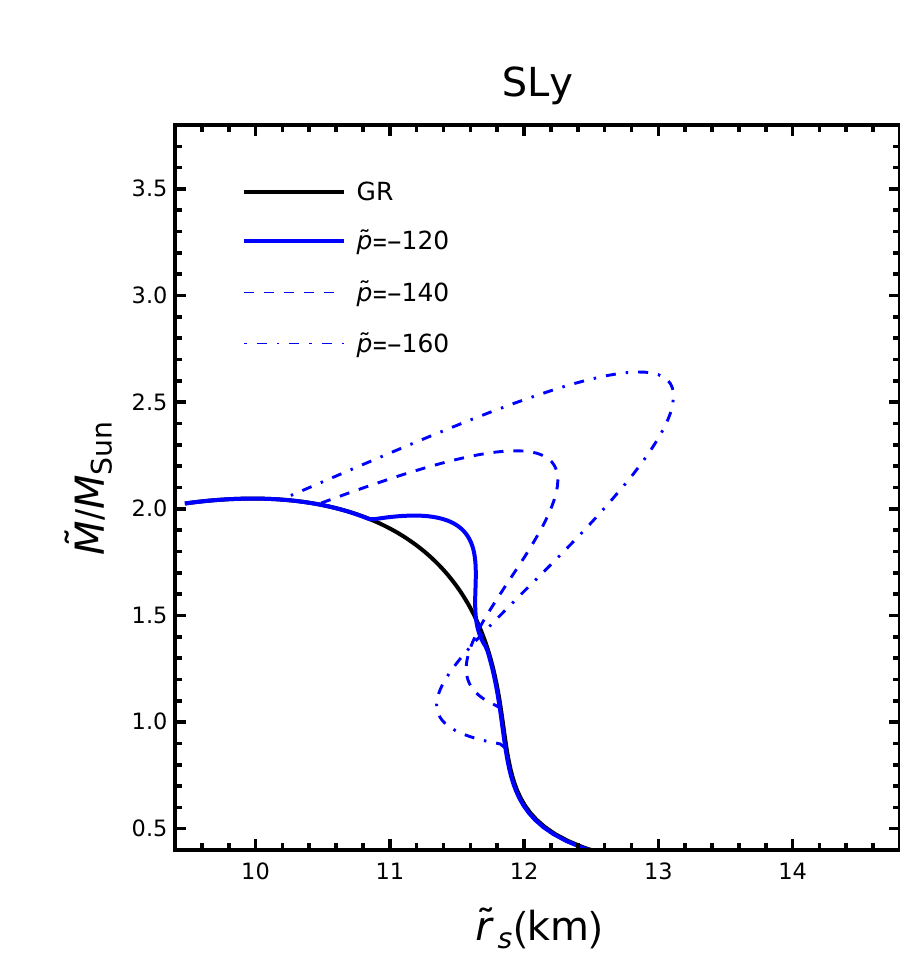} \hspace{1cm minus 0.25cm}
\includegraphics[width=0.4\textwidth, height=5cm]{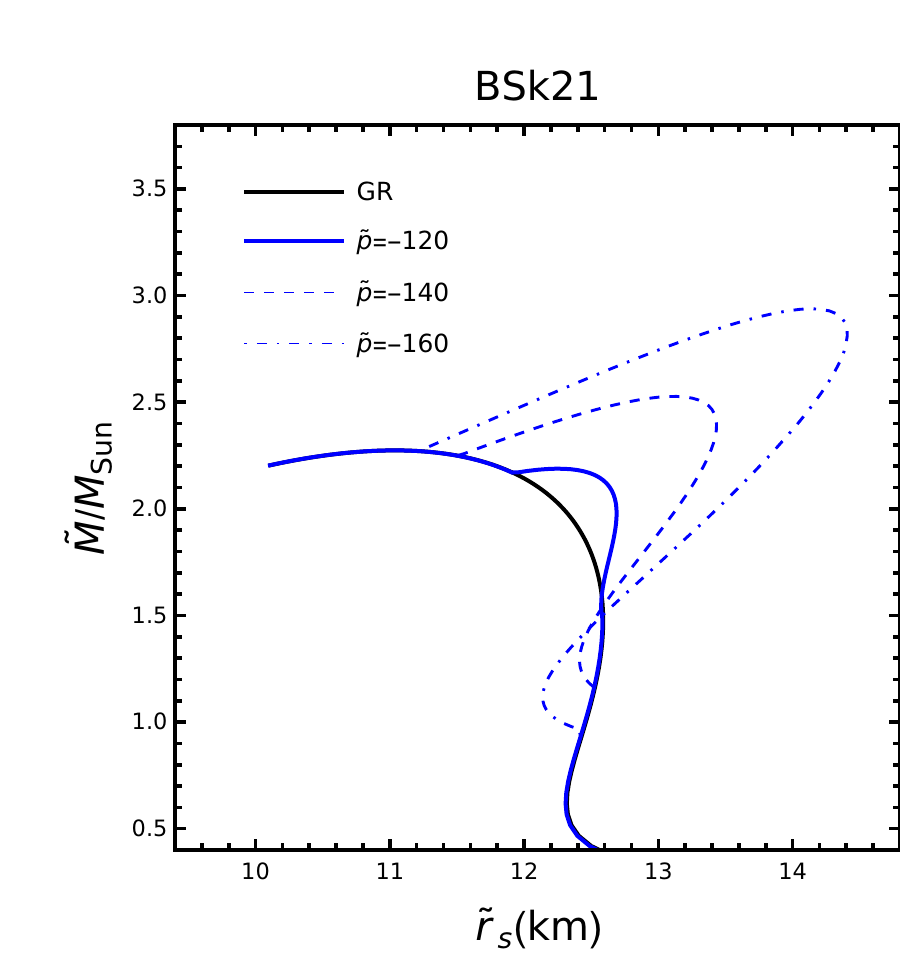} 
\caption{\small The Mass-Radius relations in Jordan frame for the equations of states: SLy (left graph) and BSk21 (right graph).}
\label{rM}
\end{figure}
 
\begin{figure}[htb]
\centering
\includegraphics[width=0.4\textwidth, height=5cm]{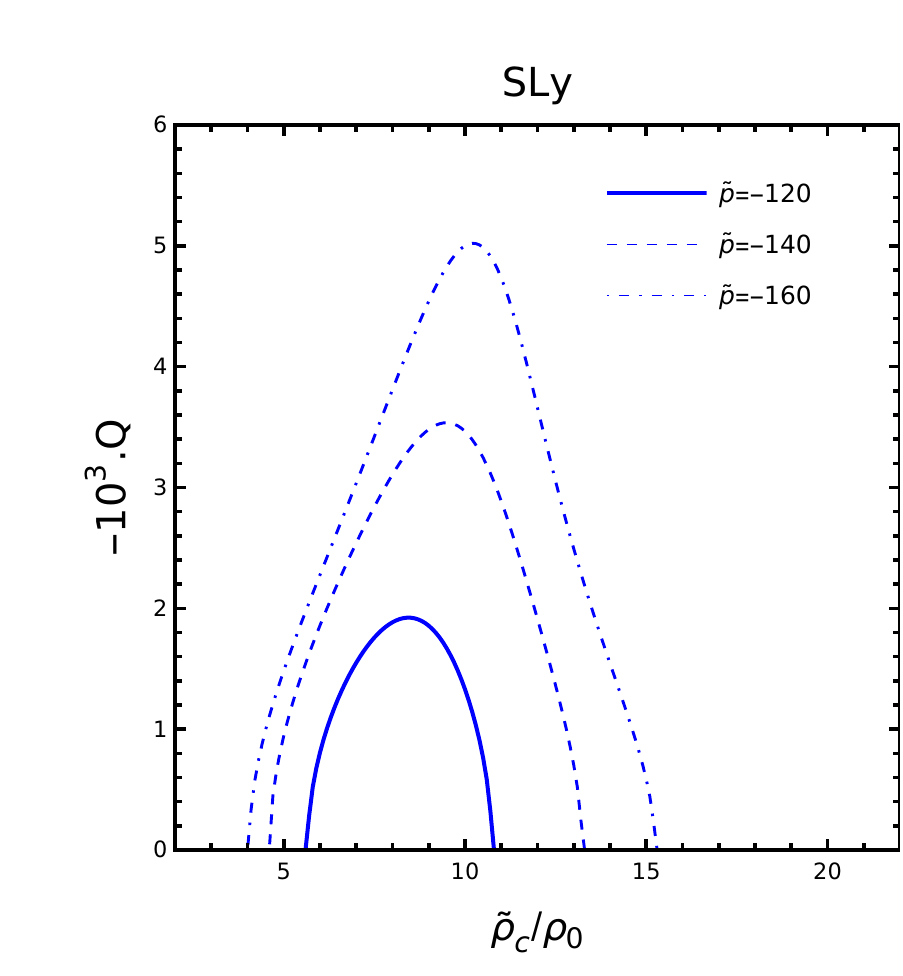} \hspace{1cm minus 0.25cm}
\includegraphics[width=0.4\textwidth, height=5cm]{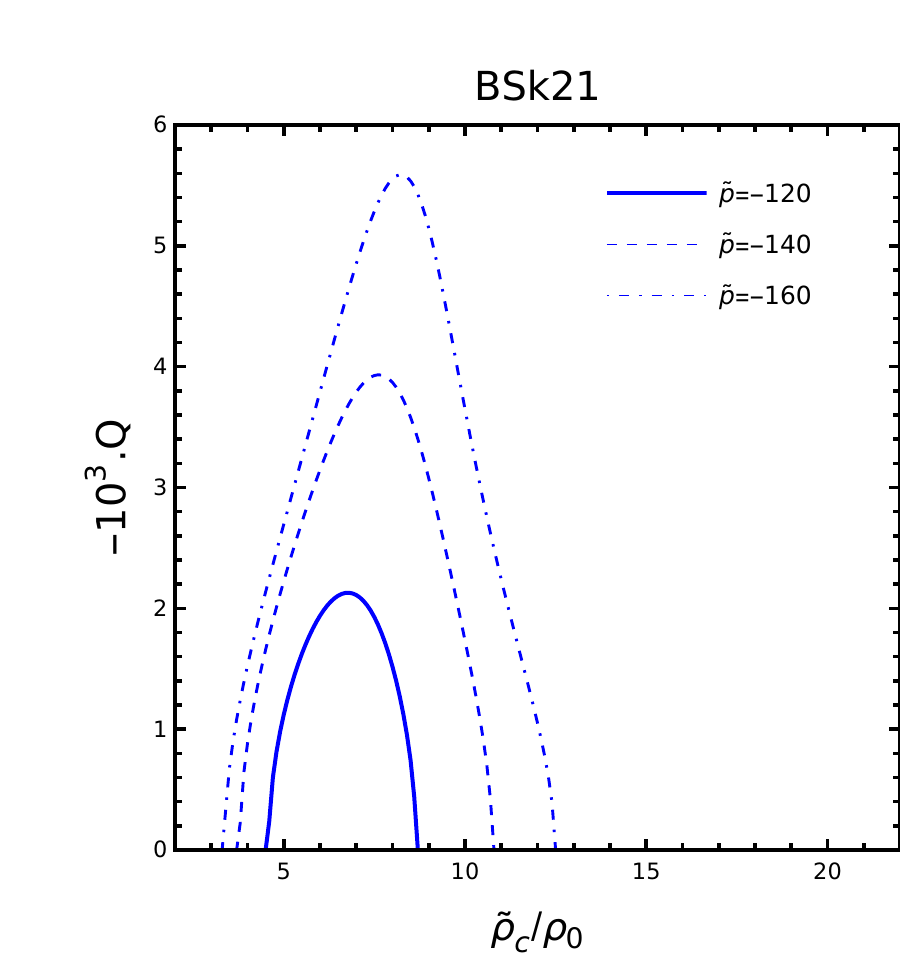} 
\caption{\small The variation of the charge $Q$ as a function of the central energy $\rho_0^c$ in Jordan frame for the equations of states:  SLy (left graph) and BSk21(right graph).}
\label{roQ}
\end{figure}

We wish to avoid singularities that appear in Eqs. (\ref{varphic}) and (\ref{Pc}) in our integration as well as  the gradient instabilities. The first problem is avoided by taking small or negative values of $\tilde{\Lambda}$ ($\tilde{\Lambda}\ll 1$ or $\tilde{\Lambda}\leq 0$). The second problem appears only for the mode $l=1$ which can be avoided for $\tilde{\Lambda}\geq 0$. Thus, the model is stable if we consider the case $\tilde{\Lambda}\ll 1$ by taking the value $\tilde{\Lambda} = 0.1$ for our numerical resolution. This value will allow us to integrate the equations for large values of $\tilde{\rho}_c$ without facing numerical instabilities from $\tilde{\rho}_c= 2 \rho_0$ to $\tilde{\rho}_c= 20 \rho_0$.

The integration is performed from the center of star $r\sim 0$ to the radius of the star $\tilde{r}_s$, defined by $\tilde{P}(r_s)=0$ where $r_s$ is the radius of the star in Einstein Frame. The relation between the two radii is given by
\begin{eqnarray}
\tilde{r}_s=\text{C}\sqrt{r_s^2+\text{D}[\varphi[r_s]]\varphi'[r_s]^2}.
\end{eqnarray} 

 Then we integrate from the surface of the star to infinity, taking into account the limit $\varphi^\infty=0$. In our paper, we use  two realistic equations of state, which are SLy and BSk21 \cite{Haensel:2004nu}. The numerical integration gives us the relation between the radius of the star and its mass. We show this relation in Fig.\ref{rM} for three values of the parameter $\tilde{p}$. Our results are similar to those found in Ref. \cite{minamitsuji2016relativistic} since we chose the same forms for the functions $\text{C}$ and $\text{D}$. We can see that the deviation from GR is significant when we increase the absolute value of $\tilde{p}$. However, the modifications, due to the scalar field, of the  mass-radius relation  are observed in a finite interval of  $\tilde{\rho}_c$ where spontaneous scalarization may arise. This interval can be determined from Fig.\ref{roQ}, where we see that the charge of the scalar field is non zero only for a finite interval of  $\tilde{\rho}_c$. For example, in Fig. \ref{roQ}, the value of $Q$ is non zero when $5.8\rho_0\leq\tilde{\rho}_c\leq 10.9\rho_0$ for the SLy EoS and $\tilde{p}=-120$. We show also in Fig. \ref{roQ} the variation of scalar field charge as a function of the central density for the EoSs SLy and BSk21. We observe that the maximum of $Q$ depends on the parameter $\tilde{p}$ and the  equation of state.

In fact, the effect of the parameter $\tilde{\Lambda}$ is not considered in our analysis because it does not have a significant effect on the relations presented in Fig. \ref{rM} and Fig. \ref{roQ}. However, if we take negative values of $\tilde{\Lambda}$, the deviation from GR and from our model will not be negligible for high negative values of $\tilde{\Lambda}$ (although we will face the gradient instability for the mode $l=1$). The results in our model are close to the purely conformal transformation, which is due to the small value of $\tilde{\Lambda}$. We note that in our model, the parameter $\tilde{p}$ is not constrained by binary-pulsar observations \cite{Freire:2012mg} since the disformal function does not vanish.

\begin{figure}[htb]
\centering
\includegraphics[width=0.4\textwidth, height=5cm]{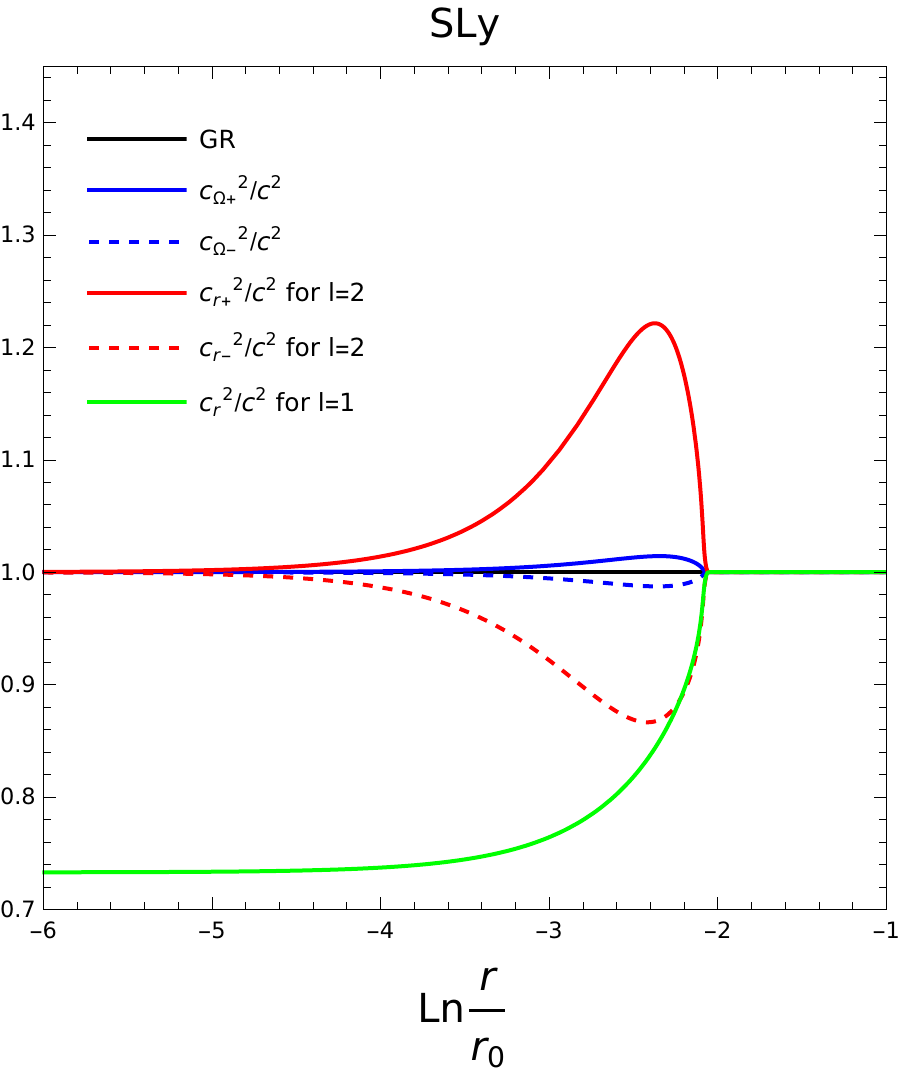} \hspace{1cm minus 0.25cm}
\includegraphics[width=0.4\textwidth, height=5cm]{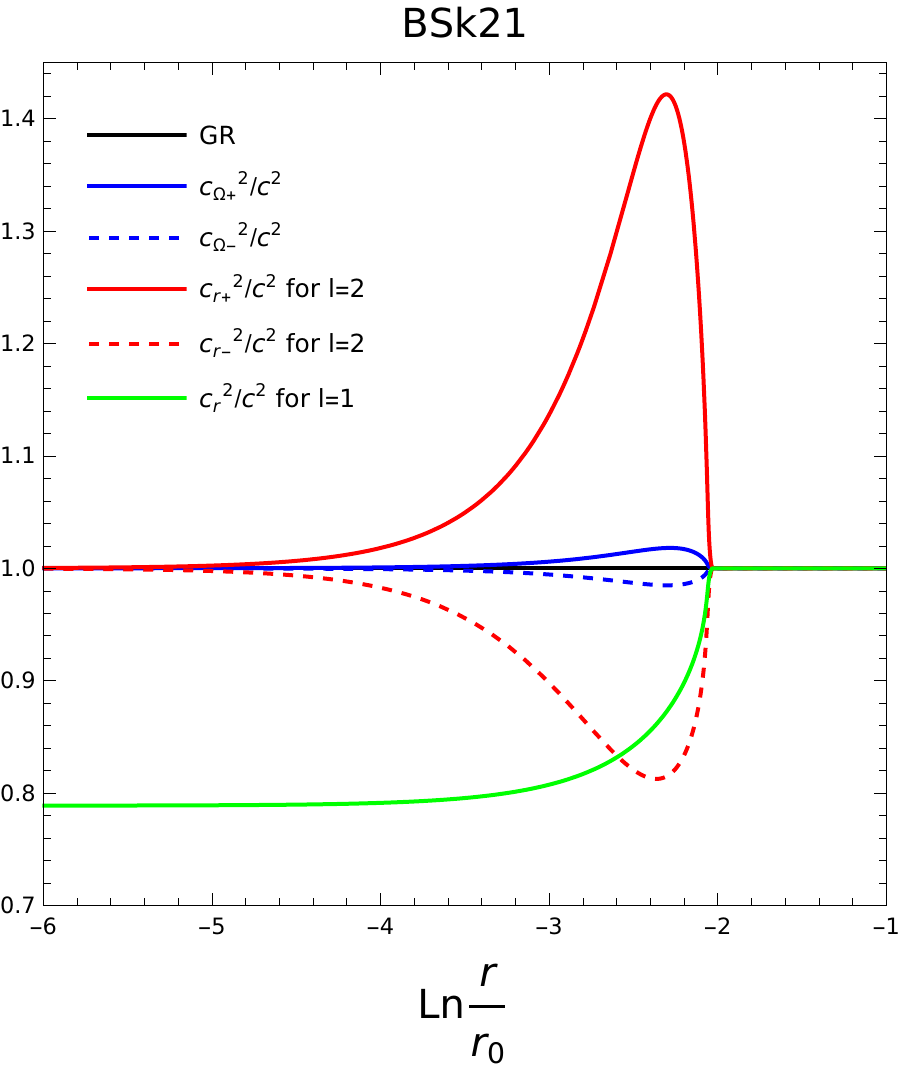} 
\caption{\small The propagation speed of the scalar field and the metric as a function of the radial coordinate $r$  for the equations of states:  SLy (left graph) and BSk21(right graph), using the central density $\tilde{\rho}_c= 12 \rho_0$.}
\label{pc}
\end{figure}

\begin{figure}[htb]
\centering
\includegraphics[width=0.4\textwidth, height=5cm]{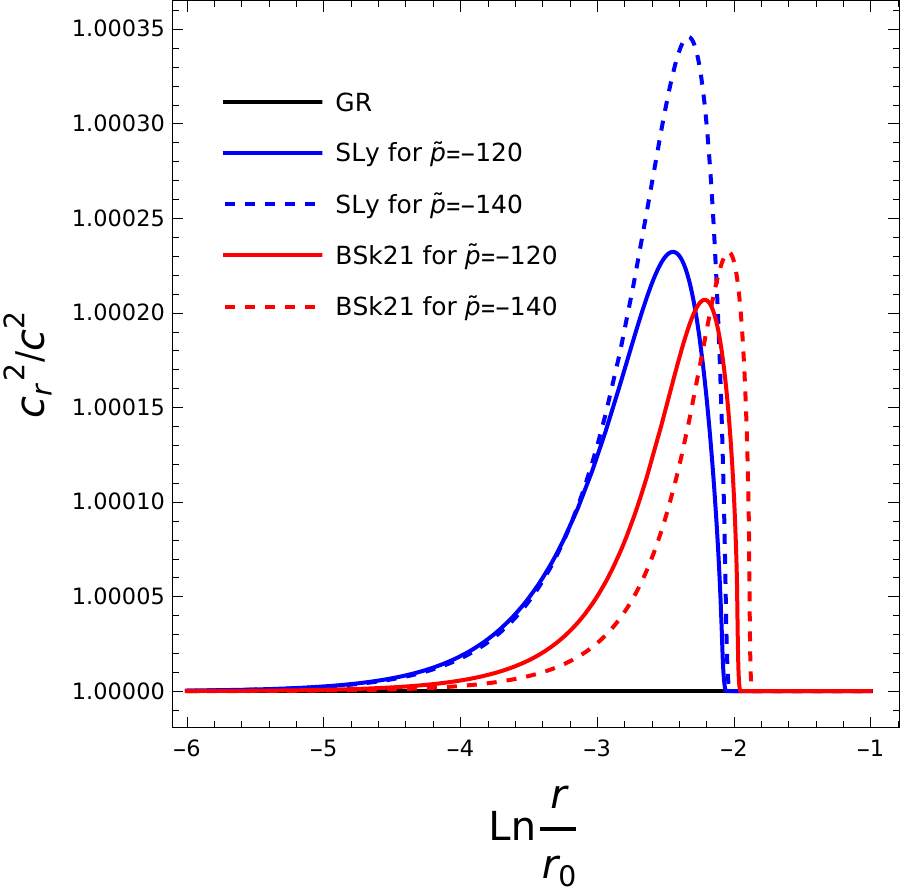}  
\caption{\small The propagation speed of the scalar field  as a function of the radial coordinate $r$  for the mode $l=0$, using the central density $\tilde{\rho}_c= 12 \rho_0$.}
\label{pc2}
\end{figure}

To confirm the stability of our model, we plot in Fig. \ref{pc} the variations of the radial and angular propagation speeds of the scalar field and the metric for the cases $l \geq 2$ (in red and blue colors) and $l=1$ (in green color) using two different  equations of state (EoSs). The results are identical to our analysis, where we observe that $c_{r\pm}$ and $c_{\Omega\pm}$ are equal to the speed of light at the center and outside the star for $l\geq 2$. In fact, the velocities $c_{r\pm}$ and $c_{\Omega\pm}$ increase (+) or decrease (-) from $c$ at $r=0$ until they reach a maximum or minimum value, depending on the value of $\tilde{p}$ and the equation of state. Then, they tend to $c$ at the radius of the star. However, for the case $l=1$, the propagation speed of $\psi$ is different from $c$ at the center, where its value can be calculated using Eq. (\ref{cr3l1}), and it is equal to $c$ outside the neutron star. In Fig. \ref{pc2}, we show that our model is also free from the gradient instability for the case $l=0$, where we observe a small deviation from General Relativity (GR) compared to the deviation in Fig. \ref{pc}.

\section{Conclusion:}

In this paper, we studied the stability of neutron stars in scalar tensor theories with a geometric metric and a physical metric related to the geometric one via a disformal transformation. In order to study the stability of neutron stars, we derived the equations of motion in a static and spherically symmetric background and then extended our study to the perturbed level by computing the conditions for the absence of ghost and gradient instabilities in the whole spacetime for the case (\ref{PC}). In addition, for a particular model described by the functions (\ref{PC}), we performed a numerical analysis aimed to show the absence of gradient instability using  two realistic equations of state.

We have seen that, for the case (\ref{PC}), a non-trivial scalar field appears with vanishing
asymptotic value, which explains the deviation from GR in the mass-radius relation. In fact, the scalar field is non zero inside the star  only for a finite interval of central densities which depends on the EoS and the value of $p$. The constant $\Lambda$ can also modify the interval, but it does not have a significant contribution. The small contribution of $\Lambda$ is due to our choice of value $\tilde{\Lambda}=0.1$ to avoid singularities at the background level and imaginary propagation speed of the scalar field at the perturbed level. Indeed,  negative values of $\Lambda$ might lead  to gradient instability for the case $l=1$ and at high central density.

Overall, our work contributes to the understanding of the behavior of neutron stars under disformal coupling, providing important insights into the stability of these astrophysical objects. Studying the stability of polar and axial perturbations is crucial because it gives us information about the solutions of the linearly perturbed equations. Due to the scalar field, which modifies the propagating speed of the perturbed  metric, matter and scalar field, it is expected that these equations and its solutions are modified with respect to those in GR.   Therefore, investigating QNMs would be interesting, and we will address this in future work.

\section*{ACKNOWLEDGEMENTS}
HB would like to express  gratitude to David Langlois for reviewing the manuscript and providing valuable comments. His insightful remarks have significantly enhanced the quality of this paper.
\newpage

\appendix
\section{Variational principle of perfect fluid in General relativity: Brief review}\label{App0}
In this appendix, we review the  thermodynamic of a single perfect fluid in general relativity. 
 The total energy density $\tilde{\rho}$ of a relativistic fluid  is the sum of its rest mass energy density  at rest $m_p c^2 N$ ( where $m_p$ is the mass of a single particle and $N$  is the baryon number)  and its internal energy density $U=\epsilon\, m_p c^2 N$, i.e.  $\tilde{\rho}=m_p c^2 \tilde{n}+ U$, where $\tilde{n}$ is the number density  in Jordan frame. The first law of thermodynamics reads
\begin{eqnarray}
\delta\epsilon = - \tilde{P} \delta\left[\frac{1}{\tilde{n}}\right]+ T \delta s,\label{depsilon}
\end{eqnarray}
where $\epsilon$, $T$ and $s$ are the specific internal energy, the temperature and the specific entropy, respectively. By defining the quantity
\begin{eqnarray}
\tilde{\mu}=\frac{\tilde{P}+\tilde{\rho}}{\tilde{n}},\label{mutermo}
\end{eqnarray}
which corresponds to the chemical potential,  one can rewrite the first law of thermodynamics as  
\begin{eqnarray}
\delta\tilde{P}= \tilde{n}\,\delta\tilde{\mu}+\tilde{n}\,T \,\delta s.
\end{eqnarray}
Hence, according to Pfaff's theorem one can express $\tilde{P}$, $\tilde{n}$ and $T$ as  functions of $\tilde{\mu}$ and $s$, i.e. 
\begin{eqnarray}
\tilde{P}=\tilde{P}(\tilde{\mu},s),\quad \tilde{n}=\tilde{n}(\tilde{\mu},s),\quad\text{and}\quad T=T(\tilde{\mu},s).\label{EOS1}
\end{eqnarray}
And thus, the energy density of the relativistic fluid is also written as a function of $\tilde{\mu}$ and $s$,
\begin{eqnarray}
\tilde{\rho}=\tilde{\rho}(\tilde{\mu},s).\label{EOS2}
\end{eqnarray}
In the case $s=constant$, the first  law of thermodynamics reduces
\begin{eqnarray}
d\tilde{P} &=& \tilde{n} \; d\tilde{\mu}.\label{dtP1}
\end{eqnarray}
Hence, $\tilde{n}$ can be defined as
\begin{eqnarray}
\tilde{n} &=  \tilde{P}_{\tilde{\mu} }= \frac{d \tilde{P}}{d \tilde{\mu}}. \label{dtP}
\end{eqnarray}
 Multiplying Eq.(\ref{mutermo}) by $\tilde{n}$  and then by varying the resulting equation, it follows that 
\begin{eqnarray}
  \tilde{\rho}_{ \tilde{\mu}}= \tilde{\mu}\tilde{P}_{ \tilde{\mu} \tilde{\mu}}. \label{dtpho}
\end{eqnarray}
Therefore, Eq.(\ref{dtP1}) can be modified, using (\ref{dtpho}), to give
\begin{eqnarray}
d\tilde{P} &=  \tilde{c}_m^2 \; d\tilde{\rho},
\end{eqnarray}
  where  $\tilde{c}_m^2\equiv \tilde{\mu}\tilde{P}_{\tilde{\mu}\tilde{\mu}} /\tilde{P}_{\tilde{\mu}}$ is the sound speed  of the fluid. By integrating, one can find an equation of state in which  the pressure is expressed as a function of the energy density 
\begin{eqnarray}
\tilde{P}=\tilde{P}(\tilde{\rho}).
\end{eqnarray}

Now, after having introduced the main thermodynamic quantities, we  present a variational principle for relativistic fluids in general relativity. We consider the action
\begin{eqnarray}
I =\int \sqrt{-\tilde{g}}\left[\frac{\kappa}{2}\tilde{R}+\tilde{P}(\tilde{\mu},s)\right]d^4 x, \label{action}
\end{eqnarray}
and define the four-dimensional  velocity vector as \cite{schutz1970perfect,schutz1977variational} 
\begin{eqnarray}
\tilde{u}_\alpha = \frac{1}{\tilde{\mu}}\left(\partial_\alpha \tilde{q} + A\partial_\alpha B+ \Theta\partial_\alpha  s\right),
\end{eqnarray}
 where $\Theta$ is a scalar field. The case of an irrotational fluid is obtained by setting  $A=const$ or $B=const$. The functions $A$ and $B$ are crucial to have the vorticity vector $\omega^\alpha \equiv (-\tilde{g})^{1/2}\epsilon^{\alpha\beta\gamma\sigma}\tilde{u}_\beta\nabla_{\sigma}\tilde{u}_\gamma$ different from zero.  The chemical potential $\tilde{\mu}$ is deduced from the normalization condition ($\tilde{u}_\alpha\tilde{u}_\beta \tilde{g}^{\alpha\beta}=-1$)
  as 
  \begin{eqnarray}
 \tilde{\mu}^2= -\tilde{g}^{\alpha\beta}\left(\partial_\alpha \tilde{q} + A\partial_\alpha B+ \Theta\partial_\alpha  s\right)\left(\partial_\beta \tilde{q} + A\partial_\beta B+ \Theta\partial_\alpha  s\right).\label{tmu22}
 \end{eqnarray}
From this result, we can vary the action (\ref{action}) with respect to $A$, $B$, $\tilde{g}_{\mu\nu}$, $\tilde{q}$, $\Theta$ and $s $ \cite{schutz1970perfect} to obtain the equations of motion. Varying (\ref{action}) with respect $A$ and $B$ gives, respectively, 
\begin{eqnarray}\label{cnstrntaB}
\tilde{u}_\alpha\partial^\alpha A=0,\quad\tilde{u}_\alpha\partial^\alpha B=0.
\end{eqnarray}
 And varying (\ref{action}) with respect to $\tilde{g}_{\mu\nu}$, $\tilde{q}$, $\Theta$ and $s $, we obtain the equations, respectively,
 \begin{eqnarray}
 &\kappa \tilde{G}_{\alpha\beta}= \tilde{T}_{\alpha\beta},\quad
  \nabla_\alpha(\tilde{n}\tilde{u}^\alpha)=0,\quad
\tilde{u}^\alpha \nabla_\alpha s=0,\quad
 &\tilde{u}^\alpha \nabla_\alpha \Theta =T.
 \end{eqnarray}
From the second and the third equations, we deduce 
\begin{eqnarray}
\tilde{u}^\alpha \nabla_\alpha \tilde{q} =-\tilde{\mu}.
\end{eqnarray}
Hence, we have obtained the field equations for gravity and matter.
\section{The expansion of $\tilde{h}$ at infinity:}\label{app.A00}
We calculate the expansion of the metric coefficient $\tilde{h}$ up to the fourth order as
\begin{eqnarray}
\tilde{h} &\sim & \text{C}^{\infty}-\frac{\text{Q} \text{C}^{\infty}_{\varphi}+2\text{M} \text{C}^{\infty}}{r}
+\frac{\frac{1}{2}\text{Q}^2 \text{C}^{\infty}_{\varphi\varphi}+3\text{M}\text{Q} \text{C}^{\infty}_{\varphi}+\text{C}^{\infty} \left(4\text{M}^2-\frac{\text{Q}^2 \text{a}}{\kappa}\right)}{r^2}\nonumber\\
&&-\frac{\frac{1}{6}\text{Q}^3 \text{C}_{\varphi\varphi\varphi}^\infty + 2\text{M}\text{Q}^2 \text{C}^{\infty}_{\varphi\varphi}+\frac{2}{3}\text{Q} \text{C}^{\infty}_{\varphi} \left(11\text{M}^2-\frac{7\text{Q}^2 \text{ a}}{4\kappa}\right)-\text{C}^{\infty} \left(\frac{5\text{M}\text{Q}^2 \text{a}}{\kappa}-8\text{M}^3\right)}{r^3}\nonumber\\
&&+\left(\frac{1}{24}\text{Q}^4 \text{C}_{\varphi\varphi\varphi\varphi}^\infty+\frac{5}{6}\text{M}\text{Q}^3 \text{C}_{\varphi\varphi\varphi}^\infty+\text{C}^{\infty}_{\varphi\varphi} \left(\frac{35\text{M}^2\text{Q}^2}{6}-\frac{2\text{Q}^4 \text{a}}{3 \kappa}\right)-\text{C}^{\infty}_{\varphi} \left(\frac{7\text{M}\text{Q}^3 \text{a}}{\kappa}-\frac{50\text{M}^3\text{Q}}{3}\right) \right.\nonumber\\
&&\left. +\text{C}^{\infty} \left(\text{Q}^2 \text{D}^\infty+\frac{16 \left(3 \kappa^2\text{M}^4-\frac{13}{4} \kappa\text{M}^2\text{Q}^2 \text{a}\right)+3\text{Q}^4 \text{a}^2}{3 \kappa^2}\right)\right)/r^4.
\end{eqnarray}
\section{Coefficients:}\label{app.A}
The coefficients that appears in $\delta^2 S$ are given by:
\begin{eqnarray}
&&a_1=\frac{1}{2} \sqrt{f} r^2 \varphi' \left(\frac{2 \text{a}}{\sqrt{h}}+\text{D}\rho \right),\quad a_2= \frac{\kappa  }{2} r \sqrt{f},\quad a_3 =- \frac{ \kappa   }{2 \sqrt{h}}\sqrt{f}   r,\nonumber\\
&& a_4=\text{C}_{\varphi } \left(-\frac{\text{D} \sqrt{f} r^2 (\text{P}+\rho ) \varphi '^2}{4 \text{C} c_m^2 \left(h+\text{D} \varphi '^2\right)}-\frac{\sqrt{f} r^2 \left(h (3 \text{P}+\rho )-2 \text{D} \rho  \varphi '^2\right)}{4 \text{C}}\right)-\frac{\text{D} r^2 f' (\text{P}+\rho ) \varphi '}{4 \sqrt{f} c_m^2 \left(h+\text{D} \varphi '^2\right)}\nonumber\\
&&-\frac{\text{D} \sqrt{f} r (\text{P}+\rho ) \varphi ' \left(r h'-4 \left(h+\text{D} \varphi '^2\right)\right)}{4 \left(h+\text{D} \varphi '^2\right)}+\frac{\sqrt{f} r^2 \text{D}_{\varphi } \left(\text{D} \rho  \left(\varphi '\right)^4+h (\text{P}+2 \rho ) \varphi '^2\right)}{4 \left(h+\text{D} \varphi '^2\right)}+\frac{\text{D} \sqrt{f} h r^2 (\text{P}+\rho ) \varphi ''}{2 \left(h+\text{D} \varphi '^2\right)},\nonumber\\
&&a_5= \kappa\sqrt{f},\quad a_6= -\frac{1}{4} \sqrt{f} \sqrt{h} \kappa  L -\frac{1}{4} \sqrt{f} \left(2 \sqrt{h} \kappa -\rho  r^2 \text{D} \varphi '^2+h r^2 (P-\rho )\right) ,\quad a_7=\frac{\kappa}{4} \sqrt{f} \sqrt{h},\nonumber\\
&&a_8=-r^2 \varphi ' \left(\sqrt{h} \rho  \text{D}+2 \text{a}\right),\quad a_9=\kappa r,\quad a_{10}=-\frac{\kappa}{2}\sqrt{h}r,\nonumber\\
&& a_{11}=\frac{1}{8} \sqrt{f} \sqrt{h} \left(2 \kappa+\frac{ \left(2 h+3 \text{D}\varphi '^2\right)}{h+\text{D} \varphi '^2}\sqrt{h} \text{P} r^2\right)-\frac{1}{8} \sqrt{f} h^2 r^2 c_m^2 (\text{P}+\rho ),\quad a_{12}=\frac{\sqrt{f} r^2 \varphi ' }{2 \sqrt{h}}\left(2 \text{a}-\frac{\text{D} h^{3/2} \text{P}}{h+\text{D} \left(\varphi '\right)^2}\right),\nonumber\\
&& a_{13}=\text{C}_{\varphi } \left(\frac{\sqrt{f} h r^2 \left(\text{D} (\text{P}-\rho ) \varphi '^2+2 h \text{P}\right)}{4 \text{C} \left(h+\text{D} \varphi '^2\right)}-\frac{3 \sqrt{f} h^2 r^2 c_m^2 (\text{P}+\rho )}{4 \text{C}}\right)+\text{D}_{\varphi } \left(\frac{\sqrt{f} h^2 r^2 c_m^2 (\text{P}+\rho ) \varphi '^2}{4 \left(h+\text{D} \varphi '^2\right)}\right.\nonumber\\
&&\left. -\frac{\sqrt{f} h \text{P} r^2 \varphi '^2}{4 \left(h+\text{D} \varphi '^2\right)}\right)+\frac{\text{D} \sqrt{f} h^2 r^2 c_m^2 (\text{P}+\rho ) \varphi ''}{2 \left(h+\text{D} \varphi '^2\right)}+\frac{\text{D} \sqrt{f} h r c_m^2 (\text{P}+\rho ) \varphi ' \left(4 \left(h+\text{D} \varphi '^2\right)-r h'\right)}{4 \left(h+\text{D} \varphi '^2\right)}\nonumber\\
&&-\frac{\text{D} h r^2 f' (\text{P}+\rho ) \varphi '}{4 \sqrt{f} \left(h+\text{D} \varphi '^2\right)},\quad a_{14} =-\frac{\kappa   \left(r f'+2 f\right)}{4 \sqrt{f}} ,\quad a_{15}=\frac{\kappa }{2} \sqrt{f} \sqrt{h},\quad a_{16}=\frac{\sqrt{h} \kappa   r^2}{4 \sqrt{f}},\nonumber\\
&& a_{17}=\sqrt{f}  r \varphi ' \left(2 \text{a}-\text{D} \sqrt{h} \text{P}\right),\quad e_2= -\frac{1}{2} \sqrt{f} r^2 \left(\frac{2 \text{a}}{\sqrt{h}}+\frac{\text{D}^2 \text{P} \varphi '^2}{h+\text{D} \varphi '^2}\right),\quad e_3 = \frac{r^2 }{2 \sqrt{f}}\left(2 \text{a} \sqrt{h}-\text{D}^2 \rho  \varphi '^2\right),\nonumber\\
&& e_4 = -\frac{\sqrt{\text{C}} \text{D} \sqrt{f} c_m^2 \varphi ' \left(\left(\varphi '\right)^2 \left(\text{a} h r^2+\text{D} \left(h^{3/2} \rho  r^2-4 \kappa \right)\right)+h^{5/2} \rho  r^2-h (h+3) \kappa \right)}{2 \kappa  r \left(h+\text{D} \varphi '^2\right)}\nonumber\\
&&-\frac{\sqrt{\text{C}} \text{D} \sqrt{f} \varphi ' \left(\text{a} r^2 \left(\varphi '\right)^2+h^{3/2} \text{P} r^2+h \kappa -\kappa \right)}{2 \kappa  r \left(h+\text{D} \varphi '^2\right)}+\frac{\sqrt{\text{C}} \sqrt{f} h c_m^2 \text{D}_{\varphi } \left(\varphi '\right)^2}{2 \left(h+\text{D} \varphi '^2\right)}+\text{C}_{\varphi } \left(\frac{\sqrt{f} h}{\sqrt{\text{C}} \left(2 \text{D} \left(\varphi '\right)^2+2 h\right)}\right.\nonumber\\
&&\left.-\frac{3 \sqrt{f} h c_m^2}{2 \sqrt{\text{C}}}\right)+\frac{\sqrt{\text{C}} \text{D} \sqrt{f} h c_m^2 \varphi ''}{h+\text{D} \varphi '^2},\quad c_1=\frac{\text{C}  \tilde{\mu}^2}{2 f^{3/2} (\text{P}+\rho ) \left(h+\text{D} \varphi '^2\right)},\quad c_2= \frac{\text{C} \tilde{\mu}^2}{2 \sqrt{f} r^2 (\text{P}+\rho )}=-\frac{c_3}{f c_m^2}\nonumber\\
&&c_4=-\frac{\text{C}  \tilde{\mu}^2 c_m^2}{r^2 (\text{P}+\rho )},\quad c_5= -\frac{\text{C} \tilde{\mu}^2 c_m^2}{2 \sqrt{f} r^2 (\text{P}+\rho )},\quad f_1 =-\frac{\sqrt{f}}{2\sqrt{h}}f_2=f_3 \frac{\sqrt{f}}{h c_m^2}= -\frac{1}{2} \sqrt{\text{C}}\tilde{\mu} \sqrt{f},
\end{eqnarray}
where 
\begin{eqnarray}
\text{P}=\frac{\text{C}^2}{\sqrt{h+\text{D} \varphi '^2}}\tilde{P},\quad \rho=\frac{\text{C}^2}{\sqrt{h+\text{D} \varphi '^2}}\tilde{\rho},\quad c_m^2=\frac{\tilde{c}_m^2}{h+\text{D} \varphi '^2}.
\end{eqnarray}

\section{The expressions of $A_i$ and $B_i$:}\label{app.B}
The expressions of $A_i$ are given:
\begin{eqnarray}
A_1 &=& 4 \text{a} h \kappa  (L-2)-\frac{\text{C}_{\varphi }}{\text{C}}\left(4 \sqrt{h} \rho  r^3  \varphi ' \left(2 \text{a}+\text{D} \sqrt{h} \rho \right) \left(h+\text{D} \varphi '^2\right)\right)-\frac{\text{C}_{\varphi }^2}{\text{C}^2 L}\left(4 \rho ^2 r^4  \left(h+\text{D} \varphi '^2\right)^2\right)\nonumber\\
&& -\text{D} \sqrt{h} \rho  \varphi '^2 \left(\text{D} L r^2 \varphi '^2 \left(2 \text{a}+\text{D} \sqrt{h} \rho \right)+2 \text{a} h L r^2+\text{D} h^{3/2} L \rho  r^2+2 \text{D} \kappa  (L-2)\right),\\
A_2&=& -2 h \kappa  (L-2) \left(4 \text{a}+\text{D} \sqrt{h} \text{P}\right)+2\sqrt{h}\left(\frac{\text{D} h^2 \kappa  (L-2) \text{P}}{h+\text{D} \varphi '^2}+2\frac{ \text{C}_{\varphi } }{\text{C}}\rho  r^3\varphi ' \left(2 \text{a} \text{D} \varphi '^2+2 \text{a} h\right.\right.\nonumber\\
&&\left.\left. -\text{D} h^{3/2} \text{P}\right) +\text{D} \varphi '^2 \left(\text{a} \text{D} L \rho  r^2 \varphi '^2+\text{a} h L r^2 (\rho -\text{P})+\text{D} \rho  \left(\kappa  (L-2)-h^{3/2} L \text{P} r^2\right)\right)\right),\\
A_3 &=& h \left(\text{D} \varphi '^2 \left(h^{3/2} L \text{P} r^2 \left(2 \text{a}-\text{D} \sqrt{h} \text{P}\right)+2 \kappa  (L-2) \left(2 \text{a}+\text{D} \sqrt{h} \text{P}\right)+2 \text{a} \text{D} \sqrt{h} L \text{P} r^2 \varphi '^2\right)\right.\nonumber\\
&&\left. +4 \text{a} h \kappa  (L-2)\right)/(h+\text{D}\varphi'^2).
\end{eqnarray}
And $B_i$   are expressed as:
\begin{eqnarray}
B_1 &=& \frac{4 \text{a} \sqrt{h} \kappa +2 \text{D} h \kappa  \rho }{h+\text{D} \varphi '^2}-\text{D} \rho  r^2 \varphi '^2 \left(2 \text{a}+\text{D} \sqrt{h} \rho \right)-2 \text{D} \kappa  \rho ,\\
B_2&=& \frac{2 \text{D}^2 \kappa  (\rho -\text{P}) \varphi '^2-8 \text{a} \sqrt{h} \kappa }{h+\text{D} \varphi '^2}-2 \text{D} r^2 \varphi '^2 \left(\text{a} (\text{P}-\rho )+\text{D} \sqrt{h} \text{P} \rho \right),\\
B_3 &=& (\text{D} \text{P} \varphi '^2 \left((\text{D} r^2 \varphi '^2+h r^2 )(2 \text{a}-\text{D} \sqrt{h} \text{P})+2 \text{D} \kappa \right)+4 \text{a} \sqrt{h} \kappa)/(h+\text{D}\varphi'^2).
\end{eqnarray}
\bibliographystyle{ieeetr}
\bibliography{bibliography}

\end{document}